\documentclass[12pt]{article}
\usepackage[margin=1in]{geometry}

\usepackage{caption}
\DeclareCaptionLabelFormat{adja-page}{\hrulefill\\#1 #2 \emph{(previous page)}}

\usepackage{authblk}

\usepackage{amsmath}
\usepackage{url}
\usepackage{graphicx}
\usepackage{color}
\usepackage{soul}

\usepackage{hyperref}
\usepackage{mathtools}

\usepackage{comment}

\begin{document}

\date{}

\title{Resonators with tailored optical path \\by cascaded-mode conversions}

\author[1,2,4]{Vincent Ginis \thanks{ginis@seas.harvard.edu}}
\author[1,4]{Ileana-Cristina Benea-Chelmus}
\author[1]{Jinsheng Lu}
\author[1,3]{Marco Piccardo}
\author[1]{Federico~Capasso\thanks{capasso@seas.harvard.edu}}

\affil[1]{Harvard John A. Paulson School of Engineering and Applied Sciences, Harvard University, Cambridge, MA 02138, USA}
\affil[2]{Data Lab / Applied Physics, Vrije Universiteit Brussel}
\affil[3]{Center for Nano Science and Technology, Istituto Italiano di Tecnologia, Milan 20133, Italy}
\affil[4]{These authors contributed equally to this work.}

\maketitle

\newpage
\textbf{Optical resonators enable the generation, manipulation, and storage of electromagnetic waves~\cite{saleh2019fundamentals}. They are widely used in technology and fundamental research, in telecommunications, lasers and nonlinear optics, ultra-sensitive measurements in cavity optomechanics, and the study of light-matter interactions in the context of cavity QED~\cite{yariv2007photonics}. The physics underlying their operation is determined by the constructive interference of electromagnetic waves at specific frequencies, giving rise to the resonance spectrum~\cite{goodman2005introduction}. This well-understood mechanism causes the limitations and trade-offs of resonator design, such as the difficulty of confining waves larger than the resonator and the fixed relationship between free spectral range, modal linewidth, and the resonator's refractive index and size~\cite{bogaerts2012silicon,koshelev2020subwavelength,odit2021observation}. Here, we introduce a new class of optical resonators, generating resonances by designing the optical path through transverse mode coupling in a cascaded process created by mode-converting mirrors ~\cite{ginis2020remote}. The generalized round-trip phase condition leads to resonator characteristics that are markedly different from Fabry-Perot resonators and can be tailored over a wide range, such as the largest resonant wavelength, the free spectral range, the linewidth, and the quality factor~\cite{bogaerts2012silicon}.
We confirm the existence of these modes experimentally in an integrated waveguide cavity with mode converters coupling two transverse modes into one supermode. The resonance signature of the cascaded-mode resonator is a spectrum resulting from the coherent superposition of the coupled transverse modes. We also demonstrate a transverse mode-independent transmission through the resonator and show that its engineered spectral properties agree with theoretical predictions. Cascaded-mode resonators introduce novel properties not found in traditional resonators and provide a mechanism to overcome the existing trade-offs in the design of resonators in various application areas.}

\medskip

Optical resonators are a cornerstone of modern physics and technology~\cite{saleh2019fundamentals,yariv2007photonics,goodman2005introduction}. These optical devices have two essential functions: they provide spectral selectivity to incident light and enhance its intensity in a small volume of space~\cite{armani2003ultra}. A prime example of a device that exploits both spectral selectivity and field amplification within a resonator is the laser, in essence, an optical cavity in which an active medium and a pumping mechanism are present~\cite{maiman1960optical}. In addition, resonators with an embedded crystalline nonlinearity enable efficient frequency doubling, sum and difference frequency generation, optical parametric amplification, and optical isolation~\cite{ilchenko2004nonlinear,turner2008ultra, ji2017ultra,sounas2018broadband,koshelev2020subwavelength}. The spectral sensitivity of resonators is also widely used in chemical, biological, and thermal spectroscopy, and in optical communication networks in filters, switches, and optical delay lines~\cite{ilchenko2006optical,bogaerts2012silicon,gagliardi2014cavity}. 
Furthermore, the transfer of momentum between light and matter~\cite{kippenberg2008cavity} can be enhanced inside a cavity, a feature widely used in cavity optomechanics. 
Optical resonators also provide an ideal platform to study and control quantum mechanical interactions~\cite{vuvckovic2001design,mabuchi2002cavity,spillane2005ultrahigh,kockum2017frequency}. 
Finally, several collective effects arise in an array of coupled optical resonators, including an effective magnetic field for photons, non-reciprocal phase shifts, and topologically protected edge states, useful for unidirectional and robust guiding of light ~\cite{hafezi2011robust, fang2012realizing, lu2014topological,pal2017observing,peng2019probing}.

Motivated by this multitude of applications, various innovations have been devised to design the properties of resonators by modifying the geometry, the medium, or the reflectors of the resonators~\cite{englund2005general,levy2011harmonic,liberal2016geometry}. 
Most recently, various novel cavity implementations have been realized where nanostructured mirrors are engineered to manipulate the cavity's phase shift or stability, or generally, to enable tunability of the transmitted field~\cite{shaltout2018ultrathin,fu2019metasurface,xie2020metasurface,xie2021sequentially}.

\section*{Theory of cascaded-mode resonances}
The functionality of electromagnetic resonators can be understood from the constructive interference of waves---creating resonant modes. A crucial parameter that determines these modes is the round-trip phase $\Delta \phi$, accumulated by the field after completing one round trip in the resonator~\cite{yariv2007photonics}. Waves that pick up a round-trip phase equal to a multiple of $2\pi$ constructively interfere with themselves and become resonant modes of the resonator (Fig.~1a). In the case of a Fabry-Perot geometry, the resonance condition is then given by 
\begin{equation}
  2Ln\frac{2\pi \nu}{c} + 2\phi_{r}= 2\pi m,
\end{equation}
where $\nu$ is the frequency of light, $m$ is an integer number representing the index of the resonant modes of frequency $\nu_m$, $c$ is the speed of light in vacuum, $L$ is the length of the resonator, $n$ is the refractive index of the material inside the resonator, and $\phi_r$ is the reflection phase at the mirrors. 
This simple equation explains two essential properties of resonators: the existence of the fundamental mode and the appearance of a spectrum with only a discrete number of modes. The resulting frequency spectrum from Eq.~(1) is then given by $\nu_m = c (m -\phi_r/\pi)/(2nL)$.

Above, we ignore the properties of the mode in the transversal plane. Typically, a discrete number of orthogonal transverse modes exist for each frequency, e.g., $\mathrm{TE}_i$ and $\mathrm{TM}_i$ waves, where each transverse mode experiences a different effective index ($n_{\mathrm{eff},i})$. As a result, the resonant modes of a resonator generally consist of a superposition of spectra, corresponding to the various families of transverse modes (Fig.~1b). The spectra are given by
\begin{equation}
\nu_{i,m} = \frac{c (m - \phi_{r,i}/\pi)}{2 n_{\mathrm{eff},i} L}.
\end{equation}

We now introduce a new type of resonator based on cascaded-mode coupling. We illustrate this principle in Fig.~1c,d: upon reflection on the rightmost mode-converting mirror, an incident wave with a particular transverse mode profile is converted into another transverse mode. When this mode returns to the leftmost mirror, another mode conversion occurs upon reflection. This cascade of mode conversions can be repeated as many times as the number of transverse modes supported by the waveguide. Finally, a ``supermode'' emerges when the wave is converted back to the original configuration of the incident mode. For resonators with $N$ different transverse modes, the round-trip phase is given by (Supplementary Materials):
\begin{equation}
  \Delta \phi = k_0L \xi \sum_{i=1}^N n_{\mathrm{eff},i} + \phi_{r,\mathrm{tot}}.
\end{equation}
Here, $k_0$ equals $2\pi/\lambda_0$ with $\lambda_0$ the vacuum wavelength, $\phi_{r,\mathrm{tot}}$ is the sum of all reflection phases, and $\xi$ is the parameter that encodes whether the contributing transverse modes appear once ($\xi=1$) or twice ($\xi=2$) in the chain. The round-trip phase is thus no longer merely determined by the length of the resonator and the refractive index but also by the number of coupled transverse modes. The corresponding resonance condition is
\begin{equation}
  \nu_m = \frac{c\left[m - \phi_{r,\mathrm{tot}}/(2\pi)\right]}{L\xi\sum_{i=1}^N n_{\mathrm{eff},i}}.
\end{equation}

The free spectral range is thus set by the sum of the round-trip optical paths of the different cascaded modes $L\xi\sum_{i=1}^N n_{\mathrm{eff},i}$ rather than by $2 n_{\mathrm{eff},i} L$ as in a conventional resonator. Next, whereas traditional resonators feature an incoherent superposition of different spectra, each corresponding to a different transverse mode, cascaded-mode resonators exhibit just one superspectrum (Fig.~1c,d).

This analysis is independent of how the mode conversions are realized. For instance, in the context of transverse modes in waveguides, a mode converter can be implemented using a specific refractive index variation (blue regions in Fig.~1a-d). The last column of Fig.~1 presents a useful abstraction to visualize and study cascaded-mode resonances using directed graphs. In this picture, cascade-mode resonances appear as cyclic graphs, which allows for studying the resonators using the properties of their associated adjacency matrix. (Supplementary Materials).

Above, we only consider the round-trip phase resonance condition to get insights into the spectrum of cascaded-mode resonators. To obtain a more accurate picture of this spectrum, we need to account for both the phase and the amplitude of the different waves. The transmission spectrum of a cascaded-mode resonator, where $N$ different forward-propagating modes are coupled with each other, is given by (Supplemental Material):  
\begin{equation}
  \vec{E}_\mathrm{out} = \sum_{i=1}^N \frac{t_{\mathrm{pt}_i} \mathrm{e}^{\mathrm{i}\phi_i}}{1-r_\mathrm{rt} \mathrm{e}^{\mathrm{i}\Delta \phi}}E_\mathrm{in} \vec{u}_{\mathrm{f}_i}.
\end{equation}
Here, $r_\mathrm{rt}$, $t_{\mathrm{pt}_i}$, $\phi_{i}$, and $\vec{u}_{\mathrm{f}_i}$ are respectively the round-trip reflection coefficient, the pass-through transmission amplitude, the transmission phase, and the unit vector of the forward propagating mode $i$ (Supplemental Materials).
The first striking feature of the spectrum is the modified fundamental mode wavelength. The largest wavelength $\lambda_0$ that can be confined in a traditional resonator is approximately that for which $\lambda_0/n_{\mathrm{eff}} = 2 L$. In the case of cascaded-mode resonances the largest wavelength is given by:
\begin{equation}
  \lambda_0 = L\xi\sum_{i=1}^N n_{\mathrm{eff},i}\left(1-\frac{\mathrm{mod}(\phi_{r,\mathrm{tot}},2\pi)}{2\pi}\right)^{-1}.
\end{equation}
This wavelength can be much larger than the resonator?s dimensions if a significant number $N$ of transverse modes are coupled. Indeed, compared with a traditional resonance, a supermode acquires a larger propagation phase in combination with a larger reflection phase. Both effects contribute to a larger round-trip phase. In Fig.~2a we visualize the ratio of the largest wavelength in a cascaded-mode resonator to that in a traditional resonator as a function of the two preceding parameters. In the Supplementary Materials, we compare this mechanism with the mechanism underlying other subwavelength resonator designs~\cite{kuznetsov2016optically,hill2007lasing,koshelev2020subwavelength,shaltout2018ultrathin}. Here, it is important to note that the local refractive index remains unchanged in a cascaded-mode resonator. The confinement occurs through the cascading of transverse modes, which increases the round-trip phase, i.e., $2L n_{\mathrm{eff},i}$ is being replaced by $L\xi\Sigma_{i=1}^N n_{\mathrm{eff},i}$.

A second interesting feature of cascaded-mode resonances, in agreement with the geometrical model described above, is the modification of the free spectral range $\Delta \nu$, given by
\begin{equation}
  \Delta \nu = \frac{c}{\xi \sum_{i=1}^N n_{\mathrm{g},i} L},
  \label{eq:fsr}
\end{equation} 
where $n_{\mathrm{g},i}$ is the group index of transverse mode $i$ at frequency $\nu$.

Finally, two other crucial, spectral parameters can be engineered in a cascaded-mode resonator by controlling the round-trip phase: the linewidth $\gamma$ and the quality factor $Q$ (Fig.~2b). Unlike the previous two parameters ($\lambda_{0,\mathrm{max}}$ and $\Delta \nu$), the linewidth and the quality factor depend on the round-trip losses (Supplementary Materials).

Not only the spectral properties but also the temporal and spatial properties of these modes can be engineered by using cascaded-mode coupling. The intracavity power build-up and the intracavity power build-up time both scale proportionally to the number of coupled modes. 
While the intensity of longitudinal modes in traditional resonators exhibits a simple standing-wave profile, the intensity profile in a cascaded-mode resonator will have a more irregular profile, potentially with many different local minima and maxima.

A unique spatial property of cascaded-mode resonators is that the propagation constant of a supermode depends on the propagation direction. This phenomenon is shown in its most straightforward implementation in Fig.~1c. When a field with transverse profile of mode~1 is incident on the left side of this resonator, a cascaded mode will exist with wave vector $k = k_0n_{\mathrm{eff},1}$ propagating from left to right, and a wave vector $k = k_0n_{\mathrm{eff},2}$ propagating from right to left. Due to the distinct propagation constants in opposite directions, directional nonlinear optical effects can occur in the resonator since the phase-matching conditions may only be satisfied in one direction~\cite{boyd2020nonlinear}. The directionality could also give additional control over chiral, optomechanical, or quantum mechanical interactions inside the resonator.

A final property of cascaded-mode resonances that deserves special attention is the existence of mode-independent spectra. Indeed, different transverse modes at the input may excite the same resonance, i.e., a mode-independent resonance. As an example, in the resonators of Fig.~1c-d the transmission spectrum (third column) is the same for the two incident transverse modes (1 and 2). We show in Supplementary Materials that the different modes that excite the same resonance in a cascaded-mode resonator can be extracted from the adjacency matrix of the graph that encodes the different mode conversions in the resonator. The mode-independent behavior of cascaded-mode resonators is a unique transmission characteristic, a feature verified experimentally in Fig.~4. This is in contrast to traditional resonators, where different transverse modes exhibit different transmission spectra. Based on this property, it becomes possible to manipulate modes with different spatial profiles in an identical way using only one resonator.

\section*{Experiments}
We experimentally realize the proposed cascaded-mode resonators using the silicon-on-insulator (SOI) platform at telecom wavelengths (1550~nm). In our on-chip implementation, the cascaded modes have distinct transverse profiles $\mathrm{TE}_i$, an in-plane polarization, and propagate along waveguides rather than in free space. The SOI platform offers design flexibility in engineering the properties of the mode converters (reflection phase and magnitude), as well as the propagation properties of all modes participating in the cascade, such as their effective indices $n_{\mathrm{eff},i}$. 

The device geometry is shown in Fig.~3a,b together with scanning electron microscope (SEM) pictures of the fabricated structures. Further details are provided in the Supplementary Materials. In general, each device consists of three main optical components: input/output waveguides that couple and guide light of chosen transverse modes to and away from the mode-converting resonators; a multi-mode waveguide section of length $L_\mathrm{wg}$ in which the cascaded modes are confined; specialized corrugated Bragg reflectors located on either side of the multi-mode waveguide that reflect one transverse mode into another. While, as described theoretically above, the number of conversions in a cascaded mode is only limited by the number of available transverse modes, we restrict our experimental demonstration to cascaded-mode resonators of the type shown in Fig.~1c that couple the two distinct transverse modes $\mathrm{TE}_0$ and $\mathrm{TE}_2$. Their transverse mode profiles are shown in the inset of Fig.~3b. Consequently, the width of the waveguide in the cavity region ($w_\mathrm{wg} = 1.07~\mathrm{\mu m}$) was chosen such that it cuts off all transverse modes of a higher order than $\mathrm{TE}_2$. (See Supplementary Materials for details on the design of the individual photonic structures and their transmission/filter performance.) In addition, the grating period $\Lambda$ of the mode converters is chosen as to satisfy the phase-matching condition and provide the necessary momentum for the mode conversion to occur on the reflected wave: $2\pi/\Lambda = \Delta \beta_{12} = \beta_1 + \beta_2$, with $\beta_1 = \beta_{\mathrm{TE}_0} = n_{\mathrm{eff,TE}_0} \omega_0/c$ and $\beta_2 = \beta_{\mathrm{TE}_2} = n_{\mathrm{eff,TE}_2} \omega_0/c$ the propagation constants of the two coupled modes. This type of coupling is typically referred to as contra-directional coupling. In the Methods, we outline in more detail the strategy for designing the cascaded-mode resonators in the SOI platform.

We now demonstrate in experiments and simulations the most evident signatures of cascaded-mode resonanators: the mode-independent spectrum with modified spectral parameters. The symmetric cascaded-mode resonator of Fig.~1c provides resonant confinement to input modes that correspond to either $\mathrm{TE}_0$ or $\mathrm{TE}_2$ transverse modes and has the same transmission spectrum for either input. We confirm this computationally in Fig.~3c, where we report the simulated field profile of the same cascaded-mode resonator for the two possible inputs and find a locally enhanced field inside the resonator in both cases. Moreover, the hybrid nature of the near-infrared cascaded mode inside the resonator becomes apparent in the zoom-in of the spatial profile shown in Fig.~3d. The field profile can be decomposed into a superposition of counter-propagating $\mathrm{TE}_0$ and $\mathrm{TE}_2$ waveguide modes that exhibit, as expected, the same beating length for both inputs (marked by the white arrow). We demonstrate this property experimentally by transmission spectroscopy and contrast it with two test Fabry-Perot resonators that employ standard mirrors and provide cavity confinement to only one of $\mathrm{TE}_0$ or $\mathrm{TE}_2$ modes. The experimental results are shown for the three cases in Fig.~4a-c and Supplemental Fig.~S11: We find that cavity modes appear, as expected, for both $\mathrm{TE}_0$ and $\mathrm{TE}_2$ modes in the case of the cascaded-mode resonator only. Moreover, the experimental results are well-reproduced by our simulations. Cavity modes appear only for one of the two modes for the conventional Fabry-Perot resonators, while light is simply transmitted for the other modes. In the Methods, we describe the spectroscopic technique used in these measurements.

Next, we analyze the resonator properties of the cavity modes associated with the cascaded-mode resonators compared to the conventional Fabry-Perot modes in Fig.~4d-f. Firstly, we show in Fig.~4d that the intra-cavity modes of the cascaded-mode resonators excited by the two inputs ($\mathrm{TE}_0$ or $\mathrm{TE}_2$) coincide in frequency. We experimentally find a negligible relative deviation between the two sets of resonant wavelengths of ($\lambda_{\mathrm{TE}_0} - \lambda_{\mathrm{TE},2})/ \lambda_{\mathrm{TE}_0} \approx 4 \mathrm{ x } 10^{-5}$. Furthermore, the quality factors of the two sets are approximately equal, as shown in Fig.~4e. Finally, the group index of the cascaded modes is approximately equal to $n_\mathrm{g} = 4.3$, regardless of whether they are excited by $\mathrm{TE}_0$ or $\mathrm{TE}_2$. In contrast, the group index of the Fabry-Perot modes are equal to $n_{\mathrm{g,TE}_0} = 3.75$ and $n_{\mathrm{g,TE}_2} = 4.85$ (Fig.~4f). This result confirms once more the cascaded-mode character of the measured spectra, particularly because the group index is approximately the arithmetic mean of the group indices of the participating transverse modes, in agreement with Eq.~(7).

\section*{Discussion}

This work shows how electromagnetic resonators can be generalized to cascaded-mode resonators, where the spectrum of supermodes reflects the generalized round-trip phase condition of a cascade of different transverse modes propagating in different directions. The theory is generally valid for any cascade of orthogonal modes inside cavities of arbitrary shape and is thus not only applicable to a cascade of transverse mode profiles of an integrated waveguide~\cite{bahari2017nonreciprocal}. Indeed, for the round trip to occur after $N$ conversions, the $N+1$-th mode in the chain needs to be indistinguishable from the first, i.e., with identical frequency, temporal shape, $k$-vector, polarization, and phase profile~\cite{limonov2017fano,shiri2020hybrid, piccardo2021roadmap}. Therefore, the theory can be applied equally well for a cascade of modes with, e.g., different spin or orbital angular momenta.

The spectral, temporal, and spatial properties are no longer solely determined by the length and refractive index of the medium, but also by the number of coupled modes. This insight allows to circumvent existing trade-offs and, for example, design resonators smaller than the local wavelength. In addition, these resonators exhibit completely new properties not found in their traditional counterparts, e.g., mode-independent resonances and directionally dependent propagation properties.

We anticipate that the concept of cascaded-mode resonators will be further exploited in a broad class of technological devices and scientific experiments since the underlying principles of cascaded-mode resonances can be extended even beyond optics.

\vspace{1cm}

\textbf{Data and Code Availability}
All data and codes associated with this manuscript will be uploaded on the zenodo database prior to publication.

\medskip
\textbf{Acknowledgements}
We acknowledge support from AFOSR grants FA550-19-1-0352 and FA95550-19-1-0135. This work was performed in part at the Center for Nanoscale Systems (CNS), a member of the National Nanotechnology Coordinated Infrastructure Network (NNCI), which is supported by the National Science Foundation under NSF Award no. 1541959.

\medskip
\textbf{Author contributions}
V.G. initiated the project and conceived the concept of cascaded-mode resonators. V.G. developed the theory with inputs from I.C.B.C., J.L., M.P., and F.C.; I.C.B.C, J.L, V.G., and M.P. designed the experiment;  I.C.B.C. fabricated the devices, carried out the measurements, and analysed the experimental data; J.L. carried out the numerical simulations. All authors contributed to the analysis, discussion and
writing of the manuscript.

\medskip
\textbf{Competing interests}
A provisional patent application has been filed on the subject of this work by the President and Fellows of Harvard College.


\newgeometry{textwidth=18cm}

\begin{figure}[t!]
\captionsetup{labelformat=empty}
\centering
\includegraphics[width=18 cm]{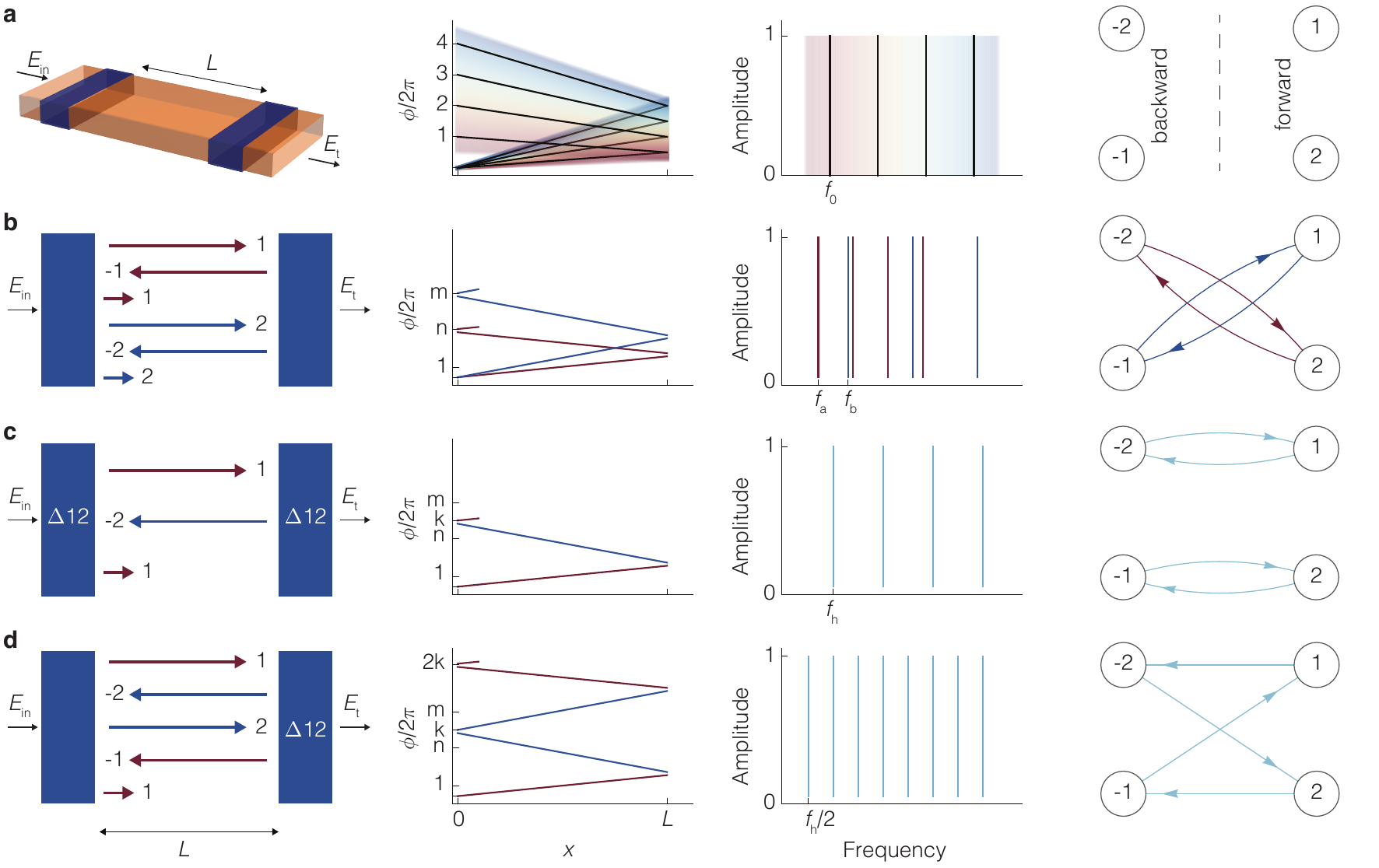} \\
\caption{}
\end{figure}

\clearpage
\begin{figure}
    \captionsetup{labelformat=adja-page}
    \ContinuedFloat
    \caption{\textbf{The operating principle underlying cascaded-mode resonances.} \textbf{a,} From left to right: First, a visualization of traditional resonator of length $L$. Second, the phase shift as a function of distance $x$ for a resonator of length $L$ for different longitudinal modes of index $m$: after a round trip $2L$ the accumulated phase is $m2\pi$. Third, the resonance spectrum, which corresponds to the frequencies for which the round-trip phase $\Delta \phi$ equals a multiple of $2\pi$. \textbf{b,} In many resonators, different transverse modes contribute to different spectra because their effective refractive indices inside the resonator differ. The different effective indices determine the different slopes of the lines in the second column, which results in spectra with different fundamental modes ($f_\mathrm{a}$, $f_\mathrm{b}$) and mode spacings in the third column. \textbf{c,} A cascaded-mode resonator (the blue regions are mode-converting mirrors) in which the two transverse modes, labeled $1$ and $2$, couple into one supermode, with fundamental frequency $f_\mathrm{h}$. The round-trip phase and the free spectral range are partly determined by the effective indices of mode 1 (red slope) and of mode 2 (blue slope).  \textbf{d,} A cascaded-mode resonator in which a supermode is created where both mode 1 and mode 2 circulate twice through the resonator before completing the round trip. There is one spectrum, with fundamental mode and free spectral range halved compared to \textbf{c}. The labels in the resonators in the first column refer to the mode conversions that take place in the blue regions: $\Delta 12$ implies that mode 1 is reflected into mode 2, and vice versa. Blue regions without label refer to traditional mirrors where each mode is reflected into itself. The fourth column illustrates the directed graph description of each resonator. In this representation, the vertices correspond to the modes and the lines between the nodes correspond to the different mode converters. The loops in these graphs identify the different resonances.}
\label{fig_Fig1}
\end{figure}

\clearpage

\begin{figure}[t!]
\centering
\includegraphics[width=9 cm]{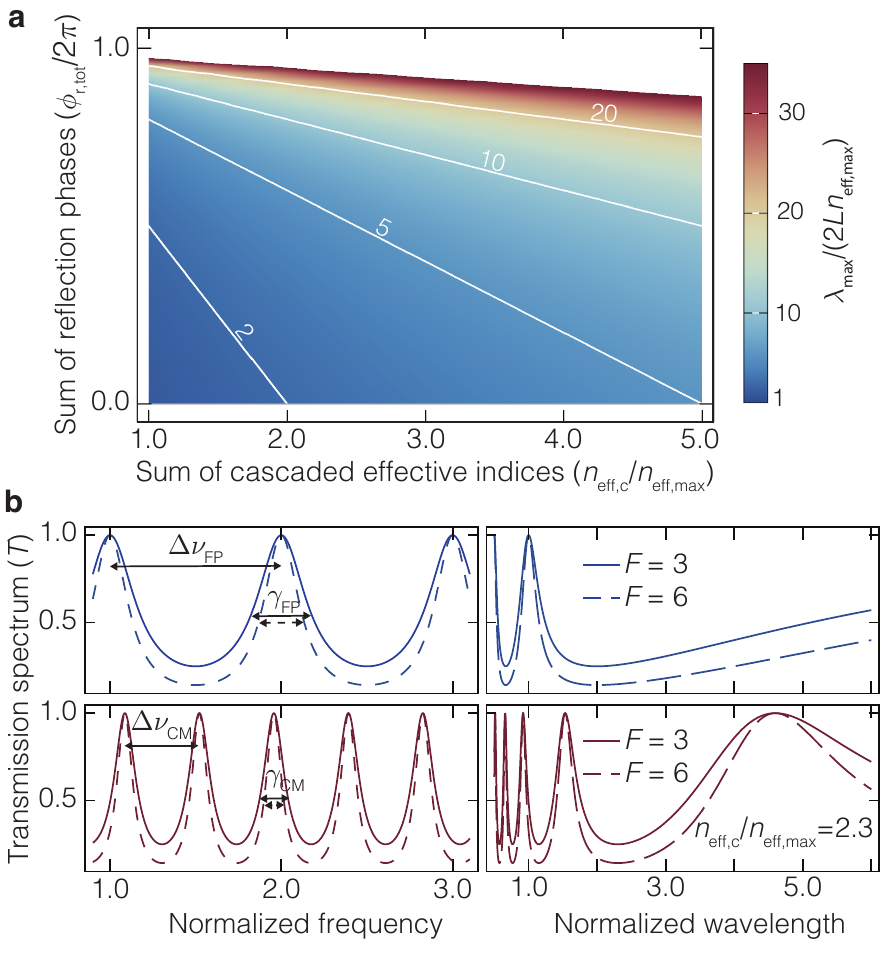} \\
\caption{\textbf{The spectral properties of cascaded-mode resonances.} \textbf{a,} The ratio of the fundamental wavelength inside a cascaded-mode resonator to the fundamental wavelength in a traditional resonator versus the sum of the cascaded effective indices $n_\mathrm{eff,c} = \xi \sum_{i=1}^N n_{\mathrm{eff},i}$, normalized by $n_\mathrm{eff,max}$, and the sum of the different reflection phases $\phi_\mathrm{r,tot}$, normalized by $2\pi$. A traditional resonator is located somewhere on the leftmost line in this parameter space ($n_\mathrm{eff,c}/n_\mathrm{eff,max}=1$). The white contour lines highlight the regions where $\lambda_\mathrm{max}/(2Ln_\mathrm{eff,max})$ equals 2, 5, 10, and 20.  \textbf{b,} The spectra of a traditional resonator compared with the spectra of a cascaded-mode resonator versus frequency (left) or wavelength (right). The blue and red spectra respectively correspond to conventional resonators and  cascaded-mode resonators with $n_\mathrm{eff,c}/n_\mathrm{eff,max} = 2.3, \phi_\mathrm{r,tot} = 1.5\pi$. The solid and dashed lines correspond to a resonator?s finesse equal to 3 and 6, respectively. Note the reduction of the free spectral range $\Delta \nu$ and linewidth $\gamma$ when several transverse modes are coupled.}
\label{fig_Fig2}
\end{figure}

\clearpage

\begin{figure}[t!]
\captionsetup{labelformat=empty}
\centering
\includegraphics[width=18 cm]{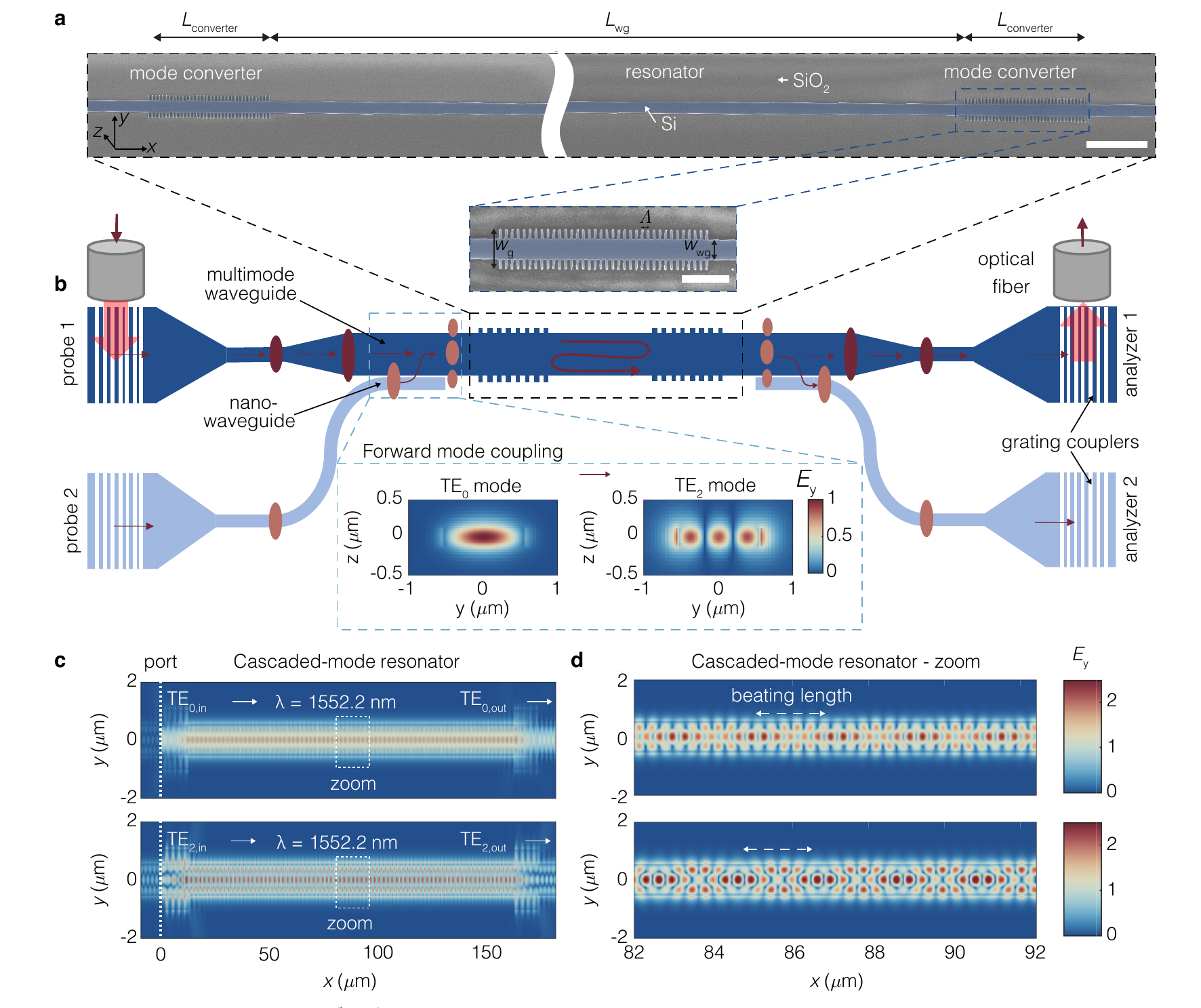} \\
\caption{}
\end{figure}

\clearpage
\begin{figure}
    \captionsetup{labelformat=adja-page}
    \ContinuedFloat
\caption{\textbf{The experimental realization of cascaded-mode resonators in integrated photonics.} \textbf{a,} SEM pictures of the cascaded-mode resonator show two mode converters that are connected via a multi-mode waveguide of width $w_\mathrm{wg}$ and length $L_\mathrm{wg}$. Multimode waveguides located before and after the resonator guide telecom light into and outside the resonator. The mode converters are realized by corrugating the silicon waveguide laterally into the shape of a rectangular grating of periodicity $\Lambda$ and width $w_\mathrm{g}$. Scalebar $= 5~\mathrm{\mu m}$ and $2~\mathrm{\mu m}$ (inset). The periodicity $\Lambda$ is chosen such that the phase-matching condition is satisfied for contra-directional coupling. The entire photonic circuit ridge is buried into a silica layer. \textbf{b,} Schematic of the device shows three different sections: 1. two input waveguides (left) that allow to probe the resonator with either $\mathrm{TE}_0$ (upper) or $\mathrm{TE}_2$ (lower), 2. the resonator region consisting of the multimode waveguide enclosed by the two mode converters, and 3. two analyzer waveguides which transmit the output of the resonator into two spatially separated locations, depending on its transverse profile $\mathrm{TE}_0$ (upper) or $\mathrm{TE}_2$ (lower). Probe 1 excites the $\mathrm{TE}_0$ mode in the top waveguide. Probe 2 excites the $\mathrm{TE}_0$ in the lower waveguide. This mode is converted into the $\mathrm{TE}_2$ mode in the top multimode waveguide prior to the resonator via the forward-mode coupler, which operates on the principle that the effective index
of the $\mathrm{TE}_0$ mode in the nano-waveguide corresponds to the effective index of the $\mathrm{TE}_2$ mode in the multimode waveguide.  Similarly analyzer 1 and analyzer 2 measure $\mathrm{TE}_0$ and $\mathrm{TE}_2$ modes, respectively. Spatially, the coupling occurs at the location where the nanowaveguide is in the immediate vicinity of the multimode waveguide. \textbf{c,} Full-wave simulations of the telecom fields inside the cascaded-mode resonator demonstrate that self-consistent solutions of the round-trip condition occur at the same input wavelength for two distinct transverse modes $\mathrm{TE}_0$ (upper) and $\mathrm{TE}_2$ (lower). \textbf{d,} Zoom into marked white region inside the resonator reveals the hybrid nature of the cascaded modes that arise as a superposition of counter-propagating $\mathrm{TE}_0$ or $\mathrm{TE}_2$ modes with a characteristic beating length that does not depend on the input probe field.}
\label{fig_Fig3}
\end{figure}

\clearpage

\begin{figure}[t!]
\centering
\includegraphics[width=18 cm]{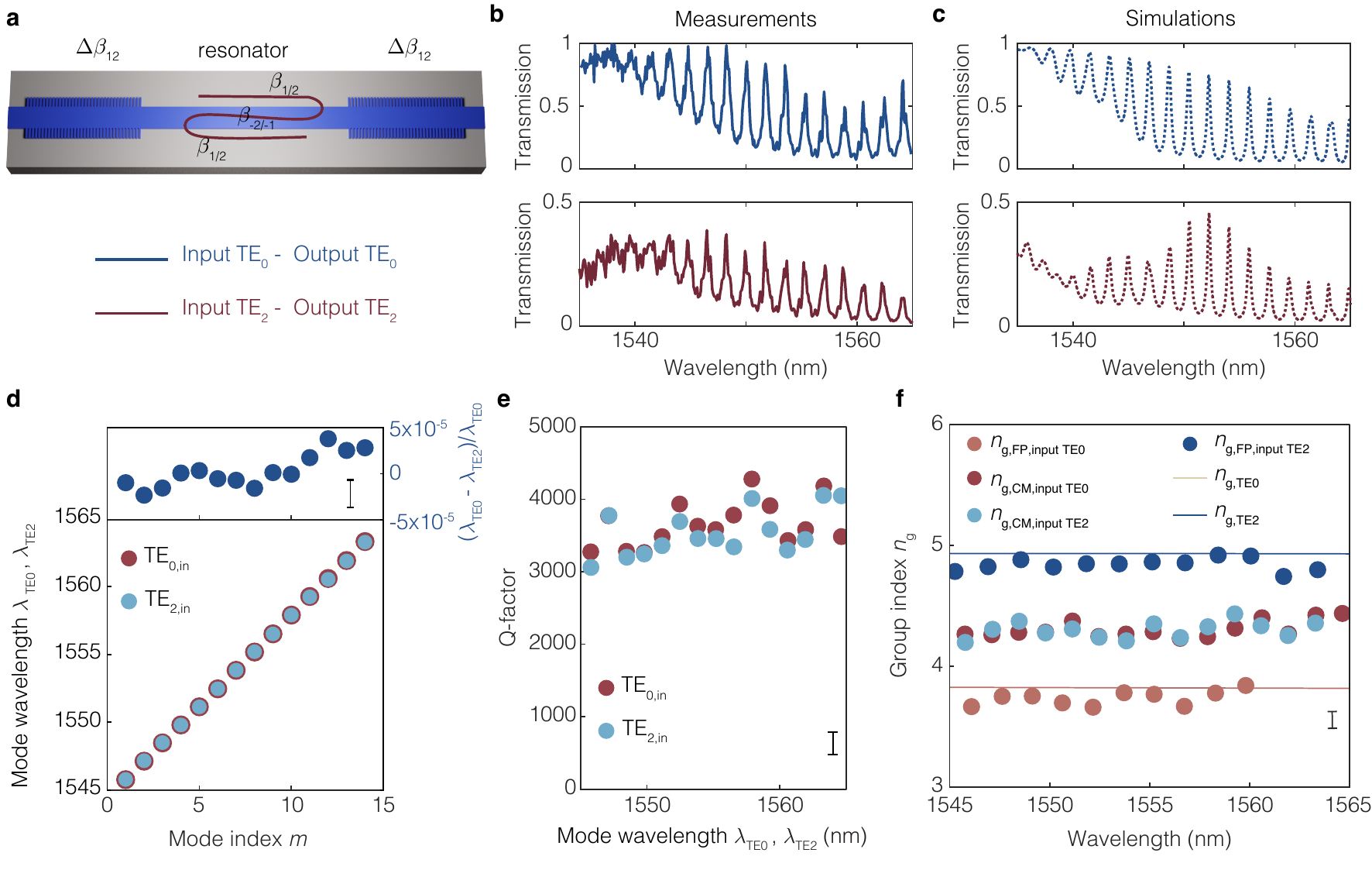} \\
\caption{}
\label{Fig4}
\end{figure}

\clearpage
\begin{figure}
    \captionsetup{labelformat=adja-page}
    \ContinuedFloat
\caption{\textbf{Transmission spectroscopy of cascaded-mode resonators.} \textbf{a,} A cascaded-mode resonator, where a reflection at both the left and the right Bragg mirror results into a conversion of the transverse mode from $\mathrm{TE}_0$ to $\mathrm{TE}_2$ and vice-versa. \textbf{b,} The measured transmission spectra of the cascaded-mode resonator exhibit resonances regardless of whether $\mathrm{TE}_0$ or $\mathrm{TE}_2$ is incident onto the resonator. \textbf{c,} Simulated transmission spectra reproduce well the measurements. More measurement and simulation results are provided in the Supplementary Material. \textbf{d-f,} Characteristic parameters of the cascaded-mode resonator extracted from the experimental transmission spectra. \textbf{d,} The mode-independent character of cascaded-mode resonators is demonstrated by the fact that the resonant wavelengths of the cavity modes coincide in frequency (and wavelength), regardless of whether the resonator is probed by a $\mathrm{TE}_0$ or $\mathrm{TE}_2$ mode. The relative experimental mismatch between their resonant wavelengths is below $4 \mathrm{x} 10^{-5}$, for all 13 considered modes, which is comparable to the experimental noise (upper panel). \textbf{e,} Also the quality factors of the cascaded-mode resonator are mode-independent, as they coincide within the experimental error, regardless of whether the cascaded modes are excited by $\mathrm{TE}_0$ or $\mathrm{TE}_2$ inputs. The experimental error is calculated from the error of the fit of the resonance peaks. \textbf{f,} In cascaded-mode resonators, the propagation constants change magnitude at each reflection off a mode-converting Bragg mirror, and the group indices of cascaded-mode (CM) resonances are equal to the arithmetic mean of conventional Fabry-Perot (FP) resonances in the same multimode waveguide. This property is confirmed experimentally, with $\mathrm{TE}_0$ modes having a group index of 3.75, $\mathrm{TE}_2$ modes having a group index of 4.85 and cascaded modes having a group index of 4.3. Full lines represent calculated group indices from full-wave simulations. Vertical bars represent error bars.}
\label{fig_Fig4}
\end{figure}

\newpage

\section*{Methods}
\subsection*{\textit{Design of the cascaded-mode resonators}}
 We adopt a rectangular grating geometry that is symmetric with respect to the center of the ridge of the waveguide, has a periodicity $\Lambda$ and a duty cycle of 40 \% as shown in Fig.~3a. The waveguide width in the corrugated area $w_\mathrm{g}$ is larger than the waveguide width in between the reflectors $w_\mathrm{wg}$. The phase-matching condition above is symmetric upon a permutation of $\beta_1$ and $\beta_2$, and the mode conversion grating is reciprocal under reflection of incident light with a transverse mode profile $\mathrm{TE}_0$ into $\mathrm{TE}_2$, and vice-versa. The selectivity of the mode converters also needs to be ensured: they need to reflect only the desired mode. We note that the grating periodicity that satisfies the contra-directional coupling condition is much shorter than the one that meets the co-directional coupling condition, in which case the converted mode would propagate in the same direction as the incident mode. Moreover, a reflection of the incident field into the same transverse mode ($\mathrm{TE}_0$ into $\mathrm{TE}_0$ or $\mathrm{TE}_2$ into $\mathrm{TE}_2$) occurs if the grating periodicity $\Lambda$ satisfies $2\pi/\Lambda = 2\beta_1$ or $2\pi/\Lambda = 2\beta_2$. To selectively satisfy only one of these phase-matching conditions and selectively convert only the desired modes, all periodicities need to be sufficiently different from each other. As a result, a design target is to engineer the effective indices of the two coupled modes to be as different as possible. This property is directly linked to the transverse dimensions of the waveguide. At the same time, we need to satisfy low propagation loss and good confinement of the $\mathrm{TE}_2$ mode to the waveguide core. As shown in Fig.~S6A, and in the mode profile of Fig.~3, a waveguide width of $w_\mathrm{wg} = 1.07~ \mathrm{\mu m}$ provides an index difference of $n_{\mathrm{eff,TE}_0} - n_{\mathrm{eff,TE}_2} = 2.751-1.944 = 0.807$. In Fig.~S10A-C we show the simulated mode conversion efficiency of the Bragg gratings used in the cascaded-mode resonator, in comparison with two test resonators which do not employ mode conversion but instead use standard Bragg mirrors that either reflect $\mathrm{TE}_0$ into $\mathrm{TE}_0$ or $\mathrm{TE}_2$ into $\mathrm{TE}_2$.

\subsection*{\textit{Spectroscopic technique}}

The spectral response of each resonator is investigated under incident light that is either prepared to be in the $\mathrm{TE}_0$ or $\mathrm{TE}_2$ transverse mode. At first, the light is directed from an optical fiber placed above the chip into low-loss single-mode waveguides via grating couplers and adiabatic tapers. The single-mode waveguides filter any undesired higher-order transverse modes that may be excited by the grating couplers. Finally, an adiabatic taper ensures a low-loss transmission of the $\mathrm{TE}_0$ mode in the top arm (Fig.~3b, dark blue) from the single-mode waveguide to the multimode waveguide that precedes the cascaded-mode resonator. Importantly, we pattern around the resonator two distinct input/output ports (Fig.~3b, dark and light blue), which allow for the preparation of the input states entering the resonators and analyzed states exiting the resonator to be either in the $\mathrm{TE}_0$ or $\mathrm{TE}_2$ transverse mode. To this end, if the light is incident in the bottom arm (Fig.~3b, light blue), we generate the $\mathrm{TE}_2$ mode from $\mathrm{TE}_0$ before the resonator using a co-directional evanescent coupler based on a nano-waveguide located next to the multimode waveguide~\cite{mohanty2017quantum}. This coupling is visualized in the inset of Fig.~3b. The $\mathrm{TE}_2$ mode is coherently excited by the $\mathrm{TE}_0$ mode. From Fig.~S6a, we find that this condition is satisfied if the nano-waveguide width equals 336 nm. The full-wave simulated field in Fig.~S7 demonstrates an efficient forward coupling with 70\% transmission at an interaction length of $L_\mathrm{int} = 60 \mathrm{\mu m}$, as used in the experiment.

\subsection*{\textit{Numerical simulations}}
Lumerical FDTD Solutions (v8.21) is used to simulate and design the mode-converting gratings and the resonant cavity composed of them. In the simulations, the thickness of the device layer of the silicon is 220 nm. The substrate is silicon oxide. A layer of silicon oxide on top with a thickness of 700 nm is applied to protect the silicon devices. The refractive index of the silicon and silicon oxide are 3.46 and 1.46, respectively. The waveguide has a width of 1100 nm, allowing for the existence of guided modes of $\mathrm{TE_0}$ and $\mathrm{TE_2}$ around a wavelength of 1550 nm. The mesh size is 25 nm. 

The period, depth, duty cycle of the gratings used for converting a forward $\mathrm{TE_0}$ mode to a backward propagating $\mathrm{TE_2}$ mode (and vice versa) is 316.5 nm, 500 nm, and $40\%$, respectively. In this case, the mode conversion efficiency reaches a maximum near 1550 nm wavelength. The sweeping range of wavelength and the number of unit cells in the grating is 1350 nm to 1750 nm and 1 to 50, respectively. We use the Mode Source Module in the Lumerical software to solve the eigenmodes in the waveguide and select the $\mathrm{TE_0}$ or $\mathrm{TE_2}$ mode as the input light source into the mode converter gratings. The transmitted and reflected electromagnetic (EM) fields are recorded after the 3D full-wave simulations. The mode expansions of the recorded EM fields are performed to obtain the transmitted and reflected power of the $\mathrm{TE_0}$ and $\mathrm{TE_2}$ modes. 
 
The resonant cavity is formed by two sets of above mentioned mode converter gratings with a cavity length of $150~\mu m$. The number of unit cells of the mode-converting mirror is chosen to be $\mathrm{N_{per}} = 36$ considering both a high mode-conversion efficiency and good bandwidth. The simulation time is set to be 72 ps which is long enough to get an accurate result. 

\subsection*{\textit{Graph representation of mode-independent resonances}}
There are generally two ways in which mode-independent resonances can appear in a cascaded-mode resonator. These two alternatives can be most easily understood by looking at the graph representation of the resonator, as shown in Extended Data Fig.~1. As described earlier, we can recognize the resonances as the loops in the directed graph. To identify the mode-independent resonances, we can write out the different loops as a sequence of the vertices.

Mode-independent resonances can then occur within one loop or between two different loops. First, within one loop, the different nodes of the sequence all contribute to the same supermode. The resonator will thus experience mode-independent resonances for each of these modes as inputs. In addition, also two different loops can give rise to the same resonance. This is the case if the sequence of one loop can be turned into the sequence of the other by inversion of the nodes and reversing the direction of the sequence. The two alternatives are shown in Extended Data Fig.~1. In Extended Data Fig.~1b, e.g., there are mode-independent resonances for input 1, input 2, and input 3: indeed, $\{1,-2,3,-3,2,-1\}$ is equivalent to $\{2,-1,1,-2,3,-3\}$ and $\{3,-3,2,-1,1,-2\}$. In the case of Extended Data Fig.~1a there are two loops in the graph: $\{1,-2,3,-4\}$ (solid lines) and $\{4,-3,2,-1\}$ (dashed lines). These loops are identical after inverting the nodes and reversing the sequence of one the loops. Incidentally, the two alternatives correspond to the parameter $\xi$ being 1, in Extended Data Fig.~1a, or 2, in Extended Data Fig.~1b.

The different modes that activate the same supermode can also be retrieved from the graph?s adjacency matrix. For example, when the adjacency matrix raised to a power $k$ has non-zero diagonal elements for some rows, then all the rows with the same diagonal element correspond to transversal modes that excite the same mode-independent resonance, as shown in Extended Data Fig.~1.

\begin{figure*}[h!]
\centering
\includegraphics[width=12 cm]{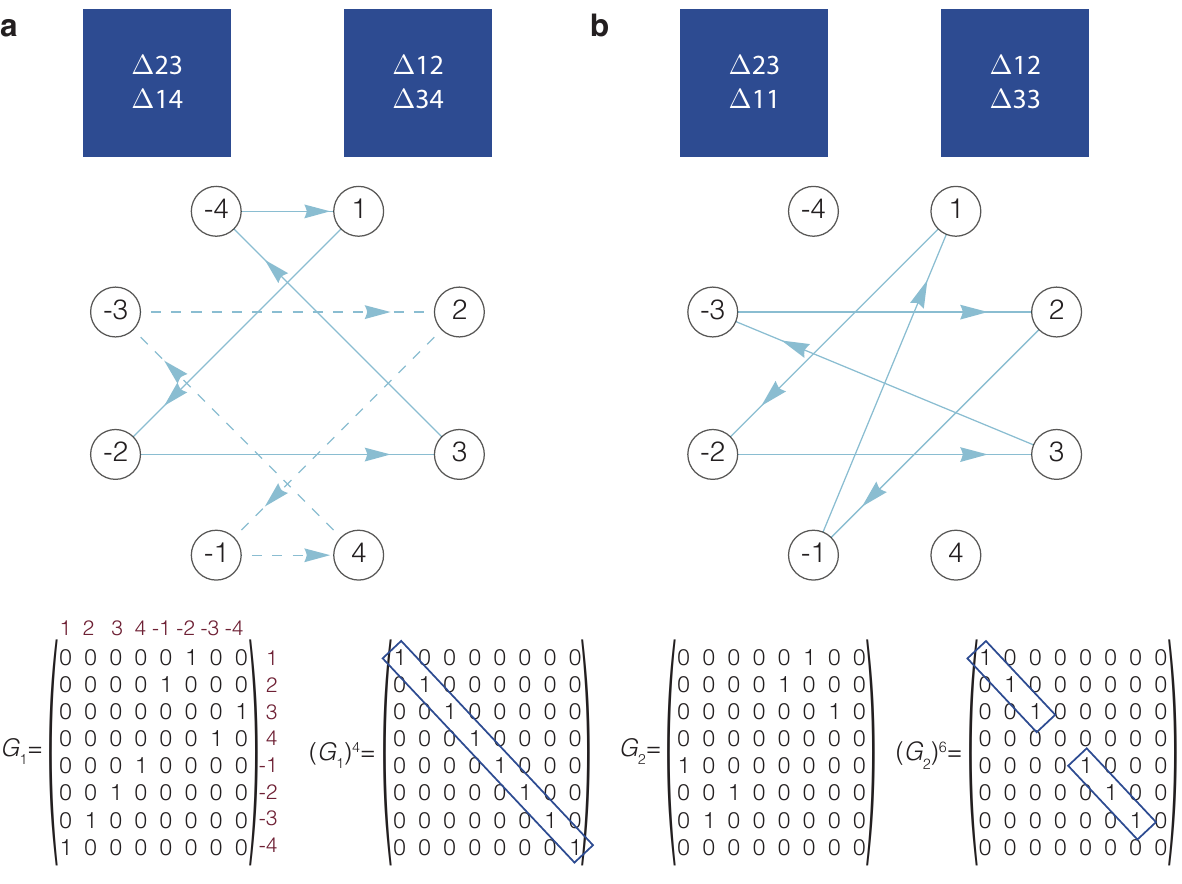} \\
 \caption*{\textbf{Extended Data Fig. 1 $\vert$ The directed graph representation to analyze mode-independent resonances in cascaded-mode resonators.} \textbf{a,} Top: a cascaded-mode resonator that couples four modes, where each mode appears only once in the resonance ($N=4, \xi = 1$). Middle: the corresponding graph of the resonator contains two cycles. The two cycles (solid lines, dashed lines) exist because of the symmetry of the mode converters: if the edge $i \to -j$ exists, then the edge $j \to -i$ also exists. Bottom: The adjacency matrix of the graph ($G_1$) and the adjacency matrix raised to the power 4, $(G_1)^4$. The adjacency matrix raised to the power $N$ is a diagonal matrix, illustrating that a node is mapped back onto itself after a path of length $N$, or a cycle of length $N$ exists in the graph. Each row with a non-zero element on the diagonal corresponds to a transversal mode that excites a mode-independent resonance. If several cascade-mode resonances of length $N$ exist in one resonator, these modes can be separated in the adjacency matrix by weighting the edges by the propagation constants of the connecting nodes. 
    \textbf{b,} A cascaded-mode resonator with three coupled modes, each appearing twice in the resonance ($N=3, \xi = 2$). The adjacency matrix ($G_2$) is diagonalized after raising the matrix to the power 6 ($=\xi N$).}
    \label{fig_ExtFig1}
\end{figure*}

\end{document}



\title{Supplementary Material for:\\
Resonators with tailored optical path \\by cascaded-mode conversions}

\author
{Vincent Ginis,$^{1, 2\ast}$ Ileana-Cristina Benea-Chelmus,$^{1\ast}$\\ Jinsheng Lu,$^{1}$ Marco Piccardo,$^{1, 3}$ and Federico Capasso$^{1}$\\
\\
\normalsize{$^{1}$Harvard John A. Paulson School of Engineering and Applied Sciences, Harvard University}\\
\normalsize{Cambridge, Massachusetts 02138, USA}\\
\normalsize{$^{2}$Data Lab / Applied Physics, Vrije Universiteit Brussel}\\
\normalsize{1050 Brussel, Belgium}\\
\normalsize{$^{3}$Center for Nano Science and Technology, Istituto Italiano di Tecnologia}\\
\normalsize{Milan 20133, Italy}\\
\normalsize{$^\ast$Authors contributed equally to this work}\\
\normalsize{E-mail: ginis@seas.harvard.edu, capasso@seas.harvard.edu.}
}

\maketitle

\date{\today}

\tableofcontents

\newpage

\section{\large{T\lowercase{heory}}}
\subsection{Maximal round-trip phase in general cascaded-mode resonators}
We calculate the maximal round-trip phase that can be obtained in a cavity of length $L$ with $N$ different transverse modes, each experiencing an effective index $n_{\mathrm{eff},i}$. A central observation in this derivation is the maximum number of times a transverse mode can be excited in one round trip. Each transverse mode can occur only two times in the round trip: once in each direction.  Indeed, each mode can only be coupled with another mode in the left and right ends of the resonator. In general, mode $i$ will be coupled to mode $j$ on the left end through the conversion $\Delta ij$, and $i$ will be coupled to $k$ on the right end through the conversion $\Delta ik$. Adding an extra conversion on either side will split the cascade of modes into sub-chains but not increase the round-trip phase of either chain.
With this knowledge, we can easily calculate that the maximum round-trip phase is obtained by having each mode occur two times - once in both directions - in one large cascade.  The total propagation round-trip phase $\Delta \phi_{p}$ that maximally can be obtained is thus given by the sum of all propagation phases:
\begin{equation}
    \Delta \phi_{p} = 2k_0L \sum_{i=1}^N n_{\mathrm{eff},i}.
\end{equation}
Additionally, at each reflection, the wave may experience a nontrivial phase shift. The total maximum round-trip phase is then given by:
\begin{equation}
    \Delta \phi = 2k_0L \sum_{i=1}^N n_{\mathrm{eff},i} + 2\sum_{i=1}^N \phi_{r,i,i+1},
\end{equation}
where $\phi_{r,i,i+1}$ is the reflection phase upon conversion from mode $i$ to $i+1$. We number the transversal modes in order of appearance in the cascade. In this notation, we implicitly assume that mode $N+1$ is the first mode again.

\subsection{The transmission spectrum of cascaded-mode resonators}
In agreement with the traditional derivation of a Fabry-Perot transmission spectrum, we now calculate the transmission spectrum of a general cascaded-mode resonator. We explicitly calculate the partially transmitted fields through the resonator, as the field inside the resonator travels back and forth. We do this in the most general case where $N$ different forward modes are coupled with each other.

In our analysis, we  number the modes $1$ to $N$. Because the cyclic nature of the mode conversions, we can rename the first node to coincide with the incident mode and subsequently rename all the other nodes in the order that they appear in the loop.

An incident field is partially transmitted through the resonator, after being transmitted through the first and second reflector and propagating through the cavity. Assuming that the product of the two transmissions through the first and second reflector is given by $t_\mathrm{pt}$, we can thus write that:
\begin{equation}
    \vec{E}_{\mathrm{out},1}=t_{\mathrm{pt}_1}\mathrm{e}^{\mathrm{i} \phi_1}E_\mathrm{in}\vec{u}_{\mathrm{f}_1}.
\end{equation}
The parameter $t_\mathrm{pt_1}$ encodes the transmission efficiency through both reflectors. The unit vector $\vec{u}_{\mathrm{f1}}$ keeps track to the vectorial nature of the field, i.e., the specific mode that is transmitted---in this case forward mode $\mathrm{f1}$.

A significant portion of the field stays inside the resonator and converts to the another mode at the second reflector. After one one backward and one forward propagation of the wave, the second partially transmitted field will be given by
\begin{equation}
    \vec{E}_{\mathrm{out},2}=t_{\mathrm{pt}_2} \mathrm{e}^{\mathrm{i}\phi_2} E_{\mathrm{in}}\vec{u}_{\mathrm{f}_2}.
\end{equation}

\begin{figure}[t!]
\centering
\includegraphics[width=16 cm]{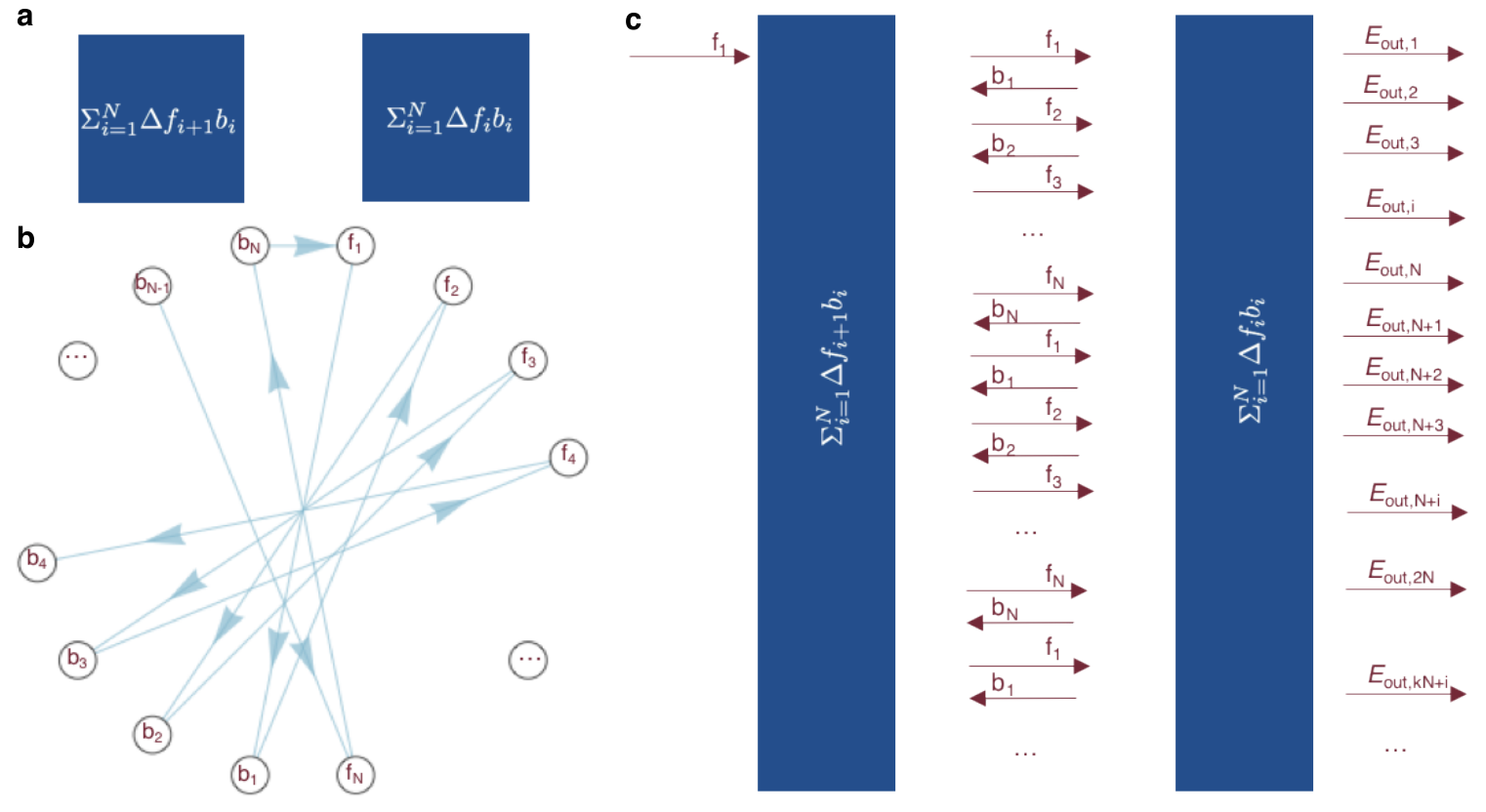} \\
\caption{\textbf{The derivation of the transmission spectrum of a general cascaded-mode resonator.} \textbf{a,}~A compact representation of the mode conversions in the resonator. The modes are renumbered in the order that they appear in the chain, where we distinguish $N$ forward propagating modes ($\mathrm{f}_i$) and $N$ backward propagating modes ($\mathrm{b}_i$). According to this nomenclature, the left and the right reflector implement $\Sigma_{i=1}^N \Delta f_{i+1} b_i$ and $\Sigma_{i=1}^N \Delta f_i b_i$, respectively. Here, we also assume $\mathrm{f}_{N+1} = \mathrm{f}_1$.  \textbf{b,}~The graph representation of the general resonator, defined in \textbf{a}. \textbf{c,}~A visualization of the partially transmitted fields at the right-hand side of the resonator. The total transmitted field is the infinite sum of all these partially transmitted fields. This sum is given by Eq.~(S12).}
\end{figure}

This picture continues until we have reached the last forward propagating mode in the system, which gives rise to the $N$th partially transmitted field:
\begin{equation}
    \vec{E}_{\mathrm{out},N}=t_{\mathrm{pt}_N} \mathrm{e}^{\mathrm{i}\phi_N} E_{\mathrm{in}}\vec{u}_{\mathrm{f}_N}.
\end{equation}

After the $N$th wave has been partially transmitted, the whole cascade has been completed and the cascade starts again. It is interesting to note that, for each mode transmitted, the following law is now fulfilled:

\begin{equation}
    \vec{E}_{\mathrm{out},kN+i} = (r_\mathrm{rt}\mathrm{e}^{\mathrm{i}\Delta \phi})^k \vec{E}_{\mathrm{out},i},
\end{equation}
where $i$ varies between $1$ and $N$ and $k$ is an integer number keeping track of the number of full round trips that have been completed.

Here, we retrieve the round-trip phase shift $\Delta \phi$, as defined in the previous section, and we introduce the parameter $r_\mathrm{rt}$:
\begin{equation}
  r_\mathrm{rt} = \prod_{i=1}^{M} r_{i,i+1},
\end{equation}
where $M$ is the total number of mode conversions in one full round trip.

The total transmitted field through the resonator, for a given incident field $\vec{E}_\mathrm{in}$ is the sum of all the partially transmitted fields, mathematically taking the sum up to infinity:
\begin{align}
    \vec{E}_\mathrm{out}&=\lim_{m\to \infty}\sum_{j=1}^m \vec{E}_{\mathrm{out},j}\\
    & = \lim_{n\to \infty}\sum_{k=0}^n\sum_{i=1}^N \vec{E}_{\mathrm{out},kN+i}\\
    & = \lim_{n\to \infty}\sum_{i=1}^N \sum_{k=0}^n \vec{E}_{\mathrm{out},kN+i}\\
   &=\sum_{i=1}^N t_\mathrm{pt_i}\vec{u}_{\mathrm{f}_i} \mathrm{e}^{\mathrm{i}\phi_i} E_{\mathrm{in}}\lim_{n\to \infty}\sum_{k=0}^n (r_\mathrm{rt} \mathrm{e}^{\mathrm{i}\Delta \phi})^k.
\end{align}
This last sum is a geometric series. In closed form it can be rewritten as $(1-r_\mathrm{rt} \mathrm{e}^{\mathrm{i}\Delta \phi})^{-1}$. The total transmitted field thus equals: 
\begin{equation}
    \vec{E}_\mathrm{out} = \sum_{i=1}^N \frac{t_{\mathrm{pt}_i} \mathrm{e}^{\mathrm{i}\phi_i}}{1-r_\mathrm{rt} \mathrm{e}^{\mathrm{i}\Delta \phi}}E_\mathrm{in} \vec{u}_{\mathrm{f}_i}
\end{equation}
In general, for a given incident field, the output will will be a sum of different forward propagating modes, whose amplitudes are given by the different $t_{\mathrm{pt},i}$. The different modes follow the same spectrum, defined by the round-trip reflection efficiency $r_\mathrm{rt}$ and the round-trip loss $\Delta \phi$.

For each of the output modes, we can calculate the transmitted intensity. Defining $t_i = E_{\mathrm{out},i}/E_\mathrm{in}$ and $\vec{t}_{\mathrm{pt},i} = t_\mathrm{pt_i}\vec{u}_i$, with $\vec{u}_i$ the unit vector of output mode $i$, we get:
\begin{align}
    T_i &= t_i^* t_i\\
    &= \frac{t_{\mathrm{pt}_i}^* t_{\mathrm{pt}_i}}{(1-r_\mathrm{rt}\mathrm{e}^{\mathrm{i}\Delta \phi})(1-r_\mathrm{rt}\mathrm{e}^{-\mathrm{i}\Delta \phi})}\\
    &= \frac{|t_{\mathrm{pt},i}|^2}{1-2r_\mathrm{rt}\cos{(\Delta \phi)}+r_\mathrm{rt}^2}\\
    &= \frac{|t_{\mathrm{pt},i}|^2}{(1-r_\mathrm{rt})^2+4r_\mathrm{rt}\sin^2{(\Delta \phi/2)}}\\
    &= \frac{\alpha_i}{1+F\sin^2{(\Delta \phi/2)}},
\end{align}
where we retrieve a generalized definition of the finesse $F = 4r_\mathrm{rt}/(1-r_\mathrm{rt})^2$ and define the normalized outgoing intensity amplitude for mode $i$: $\alpha_i = |t_{tp_i}|^2/(1-r_\mathrm{rt})^2$.

\subsection{The directed graph representation of cascaded-mode resonators}

We now discuss the relationship between the resonators, the mode converters, and the directed graphs in more detail. As shown in Fig.~S2a, cascaded-mode resonators consist of two sets of converters that convert forward propagating modes into backward propagating modes and vice-versa. One can construct the directed graph of a cascaded-mode resonator by associating each mode with a node and each mode-converter with an edge between the nodes. In doing so, it is essential to disambiguate the forward and backward propagating modes. Indeed, since there are no conversions between forward propagating modes or backward propagating modes, this procedure results in constructing a bipartite graph. The graph is directed since the mode conversions occur at one end of the resonator, and both ends of the resonator are not necessarily identical.

The general form of the adjacency matrix is shown in Fig.~S2b. The action of the converters at both ends of the resonator is visible as separate submatrices in this matrix. Here, the color of the submatrices corresponds to the color of the converters. It is now worth mentioning a subtlety about the internal symmetry of the adjacency matrix. Since both mode converters can be significantly different for one another, the adjacency matrix is not symmetric. However, the mode converters themselves are generally symmetric. For example, if a converter converts mode i to mode -j, it generally also converts mode j to mode -i. Therefore, the submatrices that implement the two converters are, in turn, symmetric matrices.

In Fig.~S2c-d, we show, by way of illustration, the graph of a specific cascaded-mode resonator where the right-hand converter consists of $\Delta 12 + \Delta 34$ and the left-hand converter consists of $\Delta 23 + \Delta 14$. The graph is shown in Fig.~S2c, where each mode conversion has a different color. In Fig.~S2d, we show the corresponding adjacency matrix where each element is circled by the color of the corresponding edge in the graph.

In Fig.~S3, we show the directed graph and the adjacency matrix of three different implementations of cascaded-mode resonators. In (a-c) we present an implementation with $N=4$ and $\xi=1$, in (d-f) an implementation with $N=4$ and $\xi=2$ and in (g-i) we have a resonator that simultaneously contains two different types of resonances: $N=4, \xi=1$ and $N=1, \xi=2$.

\begin{figure}[h!]
\centering
\includegraphics[width=12 cm]{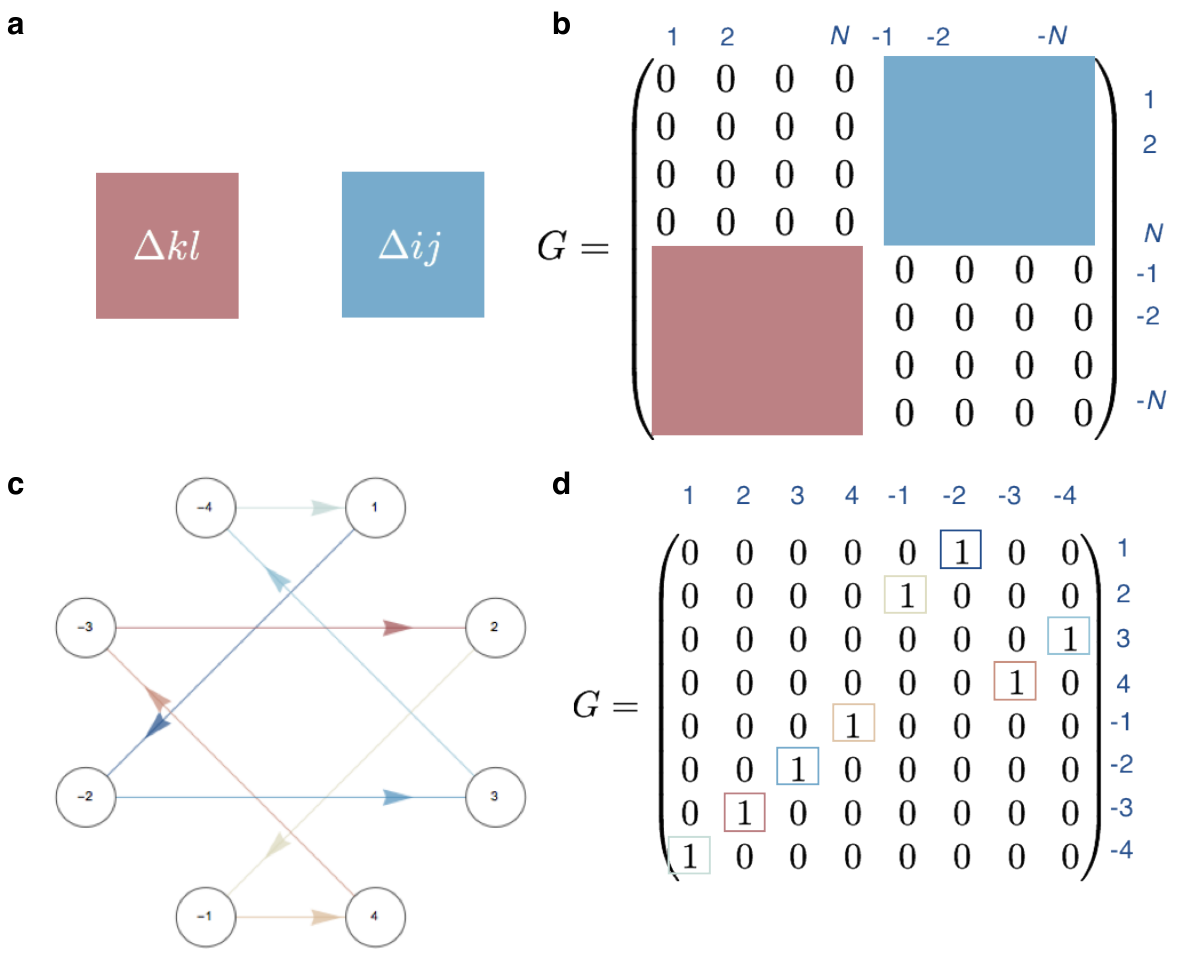} \\
\caption{\textbf{The directed graph representation of a general cascaded-mode resonator.} \textbf{a,}~A schematic of the mode conversions in a cascaded-mode resonator. \textbf{b,}~The adjacency matrix of this cascaded mode resonator contains two symmetric submatrices that correspond to the left and right mode converters. \textbf{c,}~An example of the graph in a four-mode system where the left and right converter implement $\Delta 23 + \Delta 14$ and $\Delta 12 + \Delta 34$, respectively. \textbf{d,}~The adjacency matrix of the system shown in \textbf{c}.}
\end{figure}

\begin{figure}[h!]
\centering
\includegraphics[width=12 cm]{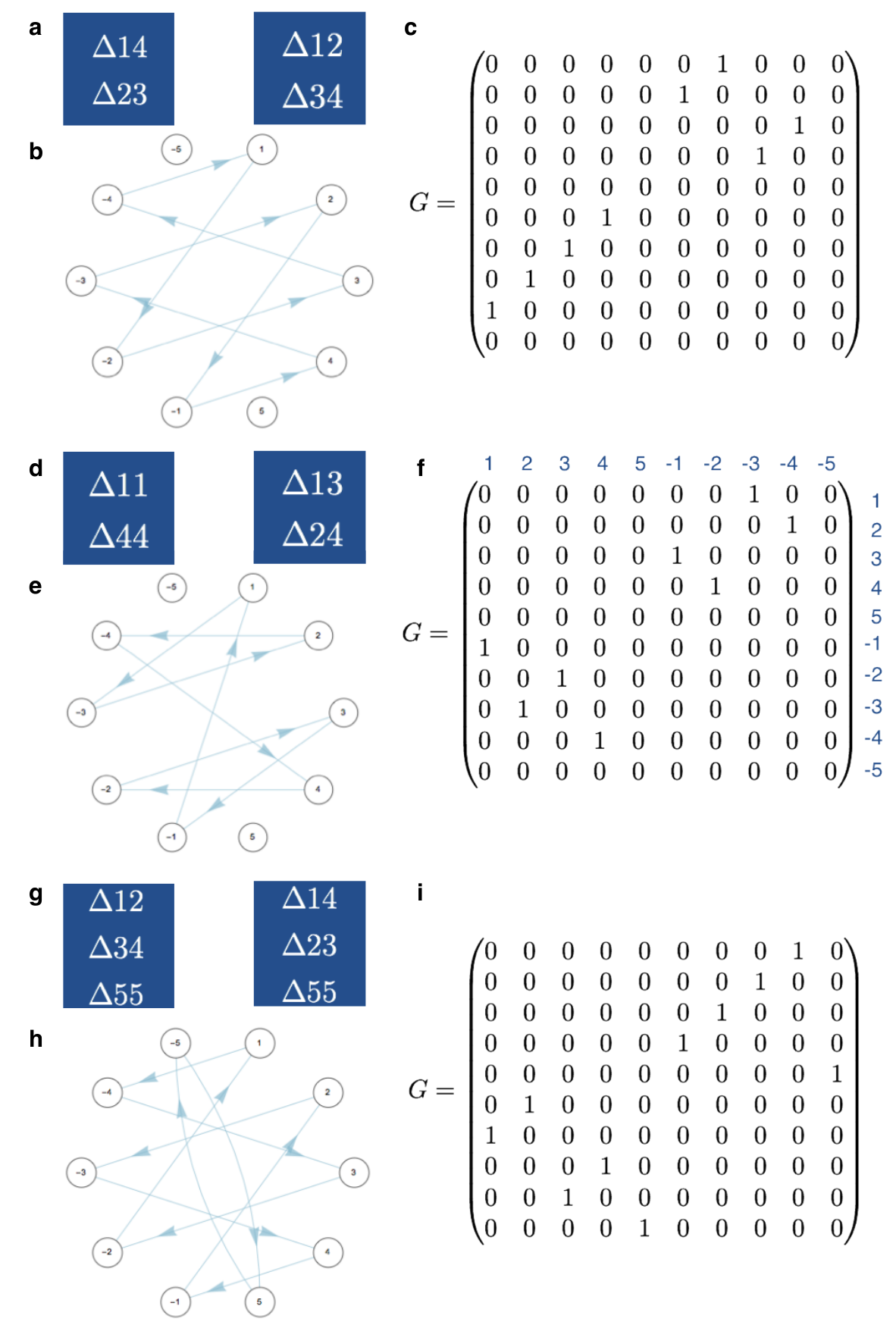} \\
\caption{\textbf{Several examples of graphs and adjacency matrices related to cascaded-mode resonators} \textbf{a-c,}~A cascaded-mode resonator where $N=4, \xi=1$.  \textbf{d-f,}~A cascaded-mode resonator where $N=4, \xi=2$. \textbf{g-i,}~A mixed-resonances cascaded-mode resonator, where two cascaded-mode resonances co-exist: $N=4,\xi=1$ and $N=1,\xi=2$.}
\end{figure}

\newpage

\subsection{Summary of scaling of spectral properties in cascaded-mode resonators}
\begin{center}
\begin{tabular}{ c c c }
 \textbf{Spectral parameter} & \textbf{traditional resonator (tr)} & \textbf{cascaded-mode resonator (CM)} \\ 
 Fundamental mode wavelength  & $\lambda_\mathrm{max,tr}$ & $\lambda_\mathrm{max,CM} = \lambda_\mathrm{max,tr}N$\\
 Free spectral range & $\Delta \nu_\mathrm{tr}$ & $\Delta \nu_\mathrm{CM}=\Delta \nu_\mathrm{tr}/N$ \\ 
 Resonance linewidth & $\Delta \gamma_\mathrm{tr}$ & $\Delta \gamma_\mathrm{CM}/N$\\ 
 Quality factor & $Q_\mathrm{tr}$ & $Q_\mathrm{CM} = Q_\mathrm{tr}N$ \\ 
 Cavity ring-down time & $\tau_\mathrm{tr}$ & $\tau_\mathrm{CM} = \tau_\mathrm{tr}N$ \\  
 Intracavity power build-up & $\kappa_\mathrm{tr} $& $\kappa_\mathrm{CM} = \kappa_\mathrm{tr}N  $  
\end{tabular}
\end{center}

\subsection{Comparison with other subwavelength resonator designs}
\begin{figure}[h!]
\centering
\includegraphics[width=6 cm]{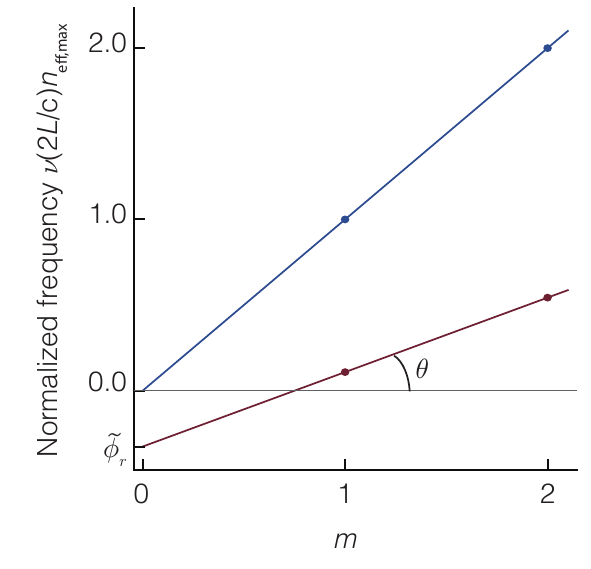} \\
 \caption{The resonance condition versus longitudinal mode number $m$, as defined in Eq.~(4) in the main paper, comparing a traditional resonator (blue) and a cascaded-mode resonator with identical $n_\mathrm{eff,max}$ (red). The sum of the coupled effective indices ($n_\mathrm{eff,c} = \xi/2 \Sigma_{i=1}^N n_{\mathrm{eff},i}$) determines the slope of the cascaded-mode spectrum ($\theta = \arctan{[(\xi/2 \Sigma_{i=1}^N n_{\mathrm{eff},i}/n_\mathrm{eff,max})^{-1}]}$. The cascaded reflection phase shift of the mode conversions ($\phi_{r,\mathrm{tot}}$) determines the intercept with the $y$ axis ($\tilde{\phi}_r = -\phi_{r,\mathrm{tot}}/(2\pi)(\xi/2 \Sigma_{i=1}^N n_{\mathrm{eff},i}/n_\mathrm{eff,max})^{-1}$. The fundamental frequency can be much lower than in a traditional resonator.}
 \label{Fig_subwavelength}
\end{figure}

The largest wavelength that can be confined in a cascaded-mode resonator is given by:
\begin{equation}
    \lambda_0 = L\xi\sum_{i=1}^N n_{\mathrm{eff},i}\left(1-\frac{\mathrm{mod}(\phi_{r,\mathrm{tot}},2\pi)}{2\pi}\right)^{-1}.
\end{equation}
Here, $\mathrm{mod}$ refers to the modulo operation, comparing the phase to the nearest multiple of $2\pi$.
The largest wavelength that can be confined effectively can be much larger than the resonator?s dimensions if a significant number $N$ of transverse modes are coupled together. This occurs through a combination of two effects, as visualized in Fig.~\ref{Fig_subwavelength}, where we show the frequency of the longitudinal modes versus mode index $m$. We compare a traditional resonator with a maximum effective index $n_\mathrm{eff,max}$ with a cascaded-mode resonator with the same $n_\mathrm{eff,max}$. The fundamental frequency of a cascaded-mode resonator is lower than that of the traditional resonator because of, on the one hand, the smaller gradient of the line connecting the resonance frequencies, given by $(\xi/2 \Sigma_{i=1}^N n_\mathrm{eff,i}/n_\mathrm{eff,max})^{-1}$, and, on the other hand, a lower intercept on the $y$ axis, determined by $- \phi_{r,\mathrm{tot}}/(2\pi)(\xi/2 \Sigma_{i=1}^N n_\mathrm{eff,i}/n_\mathrm{eff,max})^{-1}$. Physically, this can be understood as follows: compared with a traditional resonance, a supermode acquires a larger propagation phase (smaller gradient) in combination with a larger reflection phase (lower intercept). Both effects contribute to a larger round-trip phase for a fixed resonator length.

It is interesting to compare this mechanism with other resonator designs that confine waves with a free-space wavelength much larger than the dimensions of the resonators. A first example is the set of devices in which the wavelength in the resonator is adjusted, e.g., using metals or dielectrics with a large refractive index~\cite{kuznetsov2016optically,hill2007lasing,koshelev2020subwavelength}. However, there is a fundamentally different mechanism at work here. In a cascaded-mode resonator, the local refractive index remains unchanged, but the confinement occurs through the cascading of transverse modes that all combine into one mode and increase the effective phase after a round trip of that mode. In other words, $2L n_{\mathrm{eff},i}$ is being replaced by $L\xi\Sigma_{i=1}^N n_{\mathrm{eff},i}$. Second, researchers have created subwavelength resonators by adjusting the reflection phase of the mirrors, e.g., by implementing metasurface mirrors~\cite{shaltout2018ultrathin}. This mechanism is equivalent to the role played by $\phi_{r,\mathrm{tot}}$ in the above analysis. In this type of resonator, the slope of the frequency versus mode number remains along the bisector in Fig.~\ref{Fig_subwavelength}, but the line is shifted downwards.

\section{\large{M\lowercase{aterials and } \uppercase{M}\lowercase{ethods}}}

\subsection{Fabrication}

The integrated photonic circuit structures discussed in this work are fabricated using standard silicon-on-insulator~(SOI) fabrication techniques. Starting from an SOI wafer with a $2~\mathrm{\mu m}$ buried oxide layer and a $220~\mathrm{nm}$ silicon layer, we perform electron-beam lithography with an 125~keV Elionix system using ZEP520A positive resist (spin-coating at 3000~rpm, pre-exposure bake 3~min at 90$^\circ$C and 3~min at 180$^\circ$C). After exposure, we develop the photoresist in cold Oxylene for 60~s, and perform an oxygen plasma for 15~s at 40~sccm, 100~W. In a second step, the silicon layer is etched using the resist as an etch mask by single-step reactive ion etching with fluorine chemistry~($\mathrm{SF_6}$ and $\mathrm{C_4F_8}$). The buried oxide layer works as an etch stop layer. The remaining resist layer is then removed by leaving the samples overnight in Remover PG at 80$^\circ$C. A final cleaning process is performed using Piranha etch for 15~s. Finally, a 700~nm thick cladding layer of silicon dioxide is deposited via chemical vapor deposition. 

\begin{figure}[t!]
\centering
\includegraphics[width=14 cm]{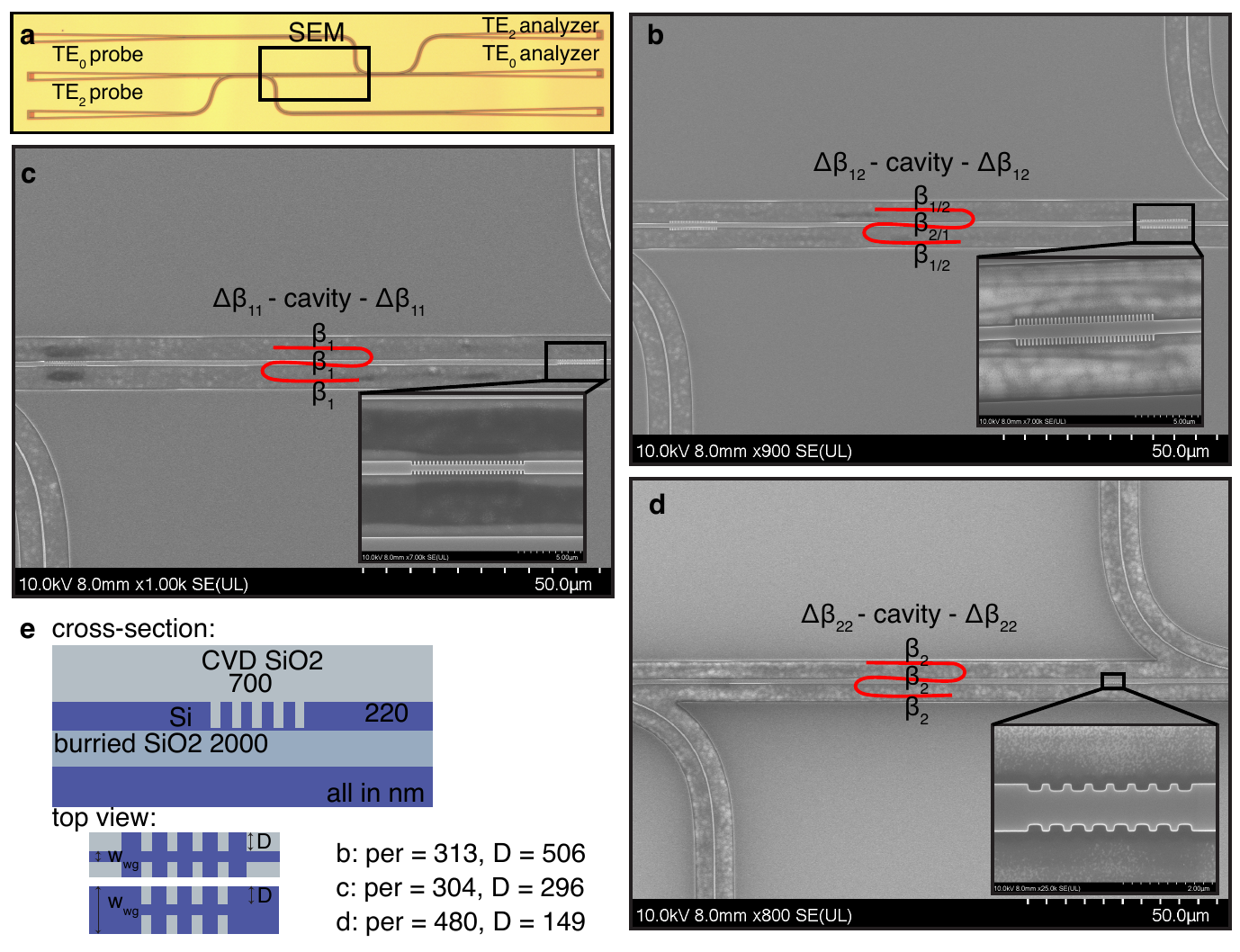} \\
\caption{\textbf{Images of the fabricated resonators.} \textbf{a,} Optical microscope picture of a fabricated resonator features three waveguide inputs on the left and three waveguide outputs on the right. From these, two sets are used to probe the resonator for input $\mathrm{TE_0}$ and $\mathrm{TE_2}$ modes and to analyze the output of the resonator also in these two transverse modes. \textbf{b-d,} SEM figures are provided for all three types of resonators we investigate in this work, and a close-up of the Bragg gratings is provided in each figure as an inset. \textbf{e,} Top-view and side-view schematics are provided for both types of Bragg reflectors. CVD = chemical vapor deposition, per = period, $\mathrm{SiO_2}$ = silicon dioxide, $\mathrm{\Delta \beta_{11}}$ = Bragg mirror that reflects selectively mode $\mathrm{TE_0}$ into $\mathrm{TE_0}$,  $\mathrm{\Delta \beta_{12}}$ = Bragg mirror that reflects mode $\mathrm{TE_0}$ into $\mathrm{TE_2}$, and vice-versa, $\mathrm{\Delta \beta_{22}}$ = Bragg mirror that reflects selectively mode $\mathrm{TE_2}$ into $\mathrm{TE_2}$.}
\label{fig_SOIresonatorSEM}
\end{figure}

A set of fabricated samples are shown in Fig.~\ref{fig_SOIresonatorSEM} by optical microscope~(a) and scanning electron microscope~(b-d). Fig~\ref{fig_SOIresonatorSEM}a illustrates one resonator structure together with waveguides that guide light of a well-defined optical mode~($\mathrm{TE_0}$ or $\mathrm{TE_2}$) into and out of the resonator, which is located in the center of the chip~(marked by the square rectangle). This allows to investigate the spectral response of the resonator for a transverse mode of choice. The SEM images in Fig~~\ref{fig_SOIresonatorSEM}b-d illustrate the marked area in the optical microscope image prior to the deposition of the silicon dioxide cladding layer for the three different types of resonators discussed in the main text: (b) the mode converting resonator where upon each reflection at the Bragg mirror, $\mathrm{TE_0}$ transverse modes are transformed into $\mathrm{TE_2}$ modes, and vice-versa; (c) a standard Fabry-Perot resonator which provides selective reflection to the $\mathrm{TE_0}$ tranverse mode, and no mode conversion occurs; and (d) a second Fabry-Perot resonator which provides selective reflection to the $\mathrm{TE_2}$ tranverse mode, and no mode conversion occurs. Each SEM figure contains an inset that shows a close-up view of the mode converter alone. For each resonator, the two Bragg gratings are identical in the cases discussed here. Fig.~\ref{fig_SOIresonatorSEM}E illustrates a top-view and a side-view schematic of the mode converters as well as the main dimensions of the chip~(per = period). The periods are equal to $\mathrm{per = 313~(b),~304~(c),~480~nm~(d)}$, the duty cycle of all gratings is $40\%$, the width of the multimode waveguide is $\mathrm{w_{wg}=1.07~\mu m}$, and the depth of the corrugations is $\mathrm{D = 506~(b),~296~(c),~149~nm~(d)}$. 

\subsection{Characterization setup}
All resonators were characterized by transmission spectroscopy using a tunable Santec TSL-550 laser, having a linewidth of 200~kHz, much below the linewidth of the resonances considered here. The polarization of the incident light is adjusted using fiber polarizers to maximize the power transmitted through the chip. Cleaved fiber probes from Lightwave at an angle of $\mathrm{10^{\circ}}$ are placed above the grating couplers located at the ends of the on-chip waveguides and couple light from the fiber to on-chip waveguides into $\mathrm{TE_0}$ mode~(in-plane polarization). The transmitted power is measured with an InGaAs photodiode with adjustable gain.

\subsection{Extraction of experimental resonator parameters}

In Fig.~5 of the main text, we report resonator parameters~(resonant wavelength, quality factor, group index) that were extracted from the experimentally measured transmission spectra. In this section, we elaborate our procedure to extract these parameters, which was applied to all data. We start by fitting Lorentzian lineshapes to each longitudinal mode and use the Lorentzian fit to determine the resonant wavelength and the Q-factor. These are reported in panels a and b. For the computation of the group index from the transmission spectra, we need an accurate estimation of the effective cavity length for each resonator. The effective cavity length $\mathrm{L_{eff} = L_{wg} +2L_{Bragg}}$ accounts for the penetration of optical fields into the Bragg reflectors~(which we summarize into an effective Bragg length $\mathrm{L_{Bragg}}$), which adds to the geometrical length of the multimode waveguide $\mathrm{L_{wg}}$. To determine $\mathrm{L_{Bragg}}$ experimentally, we fabricated in each case three separate resonators for each type of mode-coverters, each with a length of the multimode waveguide of $\mathrm{L_{wg} =100~\mu m}$, $150~\mu m$ and $200~\mu m$. Since in all three cases, $\mathrm{L_{Bragg}}$ is constant, it is possible to extract $\mathrm{L_{Bragg}}$ from the free spectral range of the longitudinal modes at different waveguide lengths. We find that $\mathrm{L_{Bragg} = 4.3~\mu m, 6.15~\mu m~and~0.5~\mu m}$ for the mode-converting grating, the $\mathrm{TE_0}$ standard Fabry Perot resonator and the $\mathrm{TE_2}$ standard Fabry Perot resonator.  By knowing the effective total cavity length, the group indices can be extracted from the spacing of adjacent longitudinal modes.

\section{\large{S\lowercase{imulations and } \uppercase{E}\lowercase{xperiments}}}

\begin{figure}[tbh]
\centering
\includegraphics[width=13 cm]{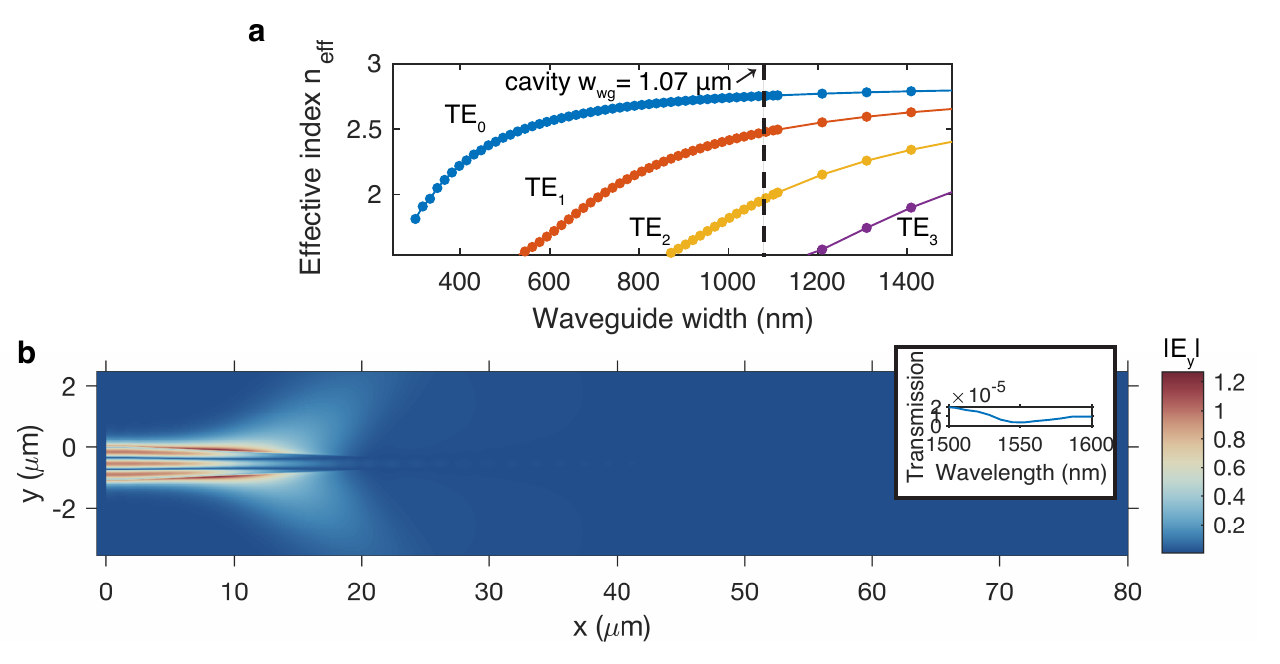} \\
\caption{\textbf{Multimode waveguide properties.} \textbf{a,} Dispersion curves of the effective indices of transverse modes $\mathrm{TE}_i$ as a function of waveguide width. \textbf{b,} Simulation of transmitted electric field of mode $\mathrm{TE}_2$ through the adiabatic taper reveals a transmission below $10^{-5}$. } 
\label{fig_SOIproperties_1}
\end{figure}

The multimode waveguide preceding the resonators, located in between the Bragg gratings and after the resonator has a width of $\mathrm{w_{wg} = 1.07~\mu m}$. This width was chosen as to maximize the difference between the effective indices of the $\mathrm{TE_0}$ and $\mathrm{TE_2}$, as to ensure a high selectivity of the three different Bragg mirrors we consider. The simulated effective indices of the modes are reported for various waveguide widths in Fig.~\ref{fig_SOIproperties_1}a. Additionally, the $\mathrm{TE_3}$ mode and all subsequent higher order modes are not supported at this waveguide width. Furthermore, we use adiabatic tapers connected to single-mode waveguides of width 440~nm to filter out any $\mathrm{TE_2}$ mode from the $\mathrm{TE_0}$ analyzer port. The efficiency of the taper and single-mode waveguide to achieve this filtering effect is simulated using finite-time-domain methods and the resulting field distribution is shown in Fig.~\ref{fig_SOIproperties_1}b. As visible from the inset, the transmission is below $10^{-5}$. 
 
\subsection{D\lowercase{esign of co-directional waveguide coupler}}

\begin{figure}[tbh]
\centering
\includegraphics[width=14 cm]{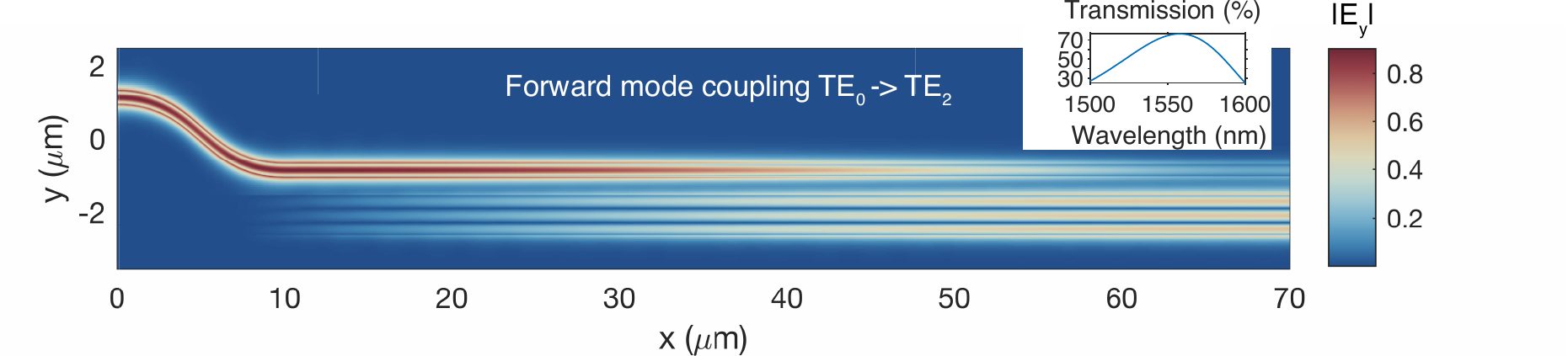} \\
\caption{\textbf{Co-directional waveguide coupler properties.} Simulation of the adiabatic forward mode coupler shows efficient energy transfer from the narrow single-mode waveguide populated with $\mathrm{TE}_0$ to the wide multimode waveguide where $\mathrm{TE}_2$ is parametrically generated. Inset: A maximal transmission of $70\%$ is achieved.} 
\label{fig_SOIproperties_2}
\end{figure}

In order to probe and analyze the resonator properties under an incident $\mathrm{TE_2}$, we design a co-directional mode converter based on the evanescent coupling of a nanowaveguide and our multimode waveguide shown in Fig.~\ref{fig_SOIproperties_2}. For the conversion to be efficient, the effective index of the $\mathrm{TE_0}$ mode in the narrow waveguide needs to match the effective index of the $\mathrm{TE_2}$ mode in the multimode waveguide. In this situation, the propagation constants of the two modes are equal and coherent injection of mode $\mathrm{TE_2}$ from $\mathrm{TE_0}$ can be ensured. Graphically, this corresponds to horizontal lines in the plot of Fig.~\ref{fig_SOIproperties_1}a, and to a width of the nanowaveguide of 335~nm. We find from full-wave simulations that a coupling length of $\mathrm{70~\mu m}$ as was used in the experiments, is sufficient to couple $70\%$ of the power into mode $\mathrm{TE_2}$ around a central wavelength of 1550~nm.

\subsection{D\lowercase{esign of contra-directional mode-converting and non-mode converting} B\lowercase{ragg gratings}}

In this section we discuss the design strategy of mode-converting and non-mode converting Bragg gratings. For these gratings to fulfill their intended purpose, several conditions need to be satisfied concomitantly: 1. all gratings provide selective and efficient reflection for a pre-selected transverse mode profile, 2. all gratings provide only minimal reflection to all other transverse mode profiles, and 3. all gratings effect only negligible co-directional mode conversion. As we anticipate above, these conditions can be fulfilled by careful choice of the effective refractive index of the transverse modes supported by the waveguide. In general, a Bragg grating provides an additional momentum that can be leveraged to achieve phase matching for a given set of modes. For contra-directional coupling of modes with propagation constants $\mathrm{\beta_1 = n_{eff,1}\frac{\omega}{c_0}}$, and $\mathrm{\beta_2 = n_{eff,2}\frac{\omega}{c_0}}$~(corresponding to mode-converting Bragg mirrors), the grating period $\mathrm{\Lambda}$ needs to be chosen such that $\mathrm{\frac{2\pi}{\Lambda} = \beta_1 + \beta_2}$. For co-directional coupling, the grating period $\mathrm{\Lambda}$ needs to be chosen such that $\mathrm{\frac{2\pi}{\Lambda} = \beta_1 - \beta_2}$. We recognize here already that the corresponding grating periods will be considerably distinct and that thereby, in general, the co-directional coupling will be negligible when contra-directional coupling is achieved. For contra-directional coupling of modes with equal propagation constants $\mathrm{\beta_1 = n_{eff,1}\frac{\omega}{c_0}}$, or $\mathrm{\beta_2 = n_{eff,2}\frac{\omega}{c_0}}$~(corresponding to mode-converting Bragg mirrors), the grating period $\mathrm{\Lambda}$ needs to be chosen such that $\mathrm{\frac{2\pi}{\Lambda} = 2\beta_1~or~2\beta_2}$~(corresponding to standard non-mode converting Bragg mirrors). Having chosen the two modes $\mathrm{TE_0}$ and $\mathrm{TE_2}$ to have maximally different effective refractive indices, it is possible to choose a grating period that achieves only mode-converting reflection and negligible standard reflection, or vice-versa. 

\begin{figure}[tbh]
\centering
\includegraphics[width=14 cm]{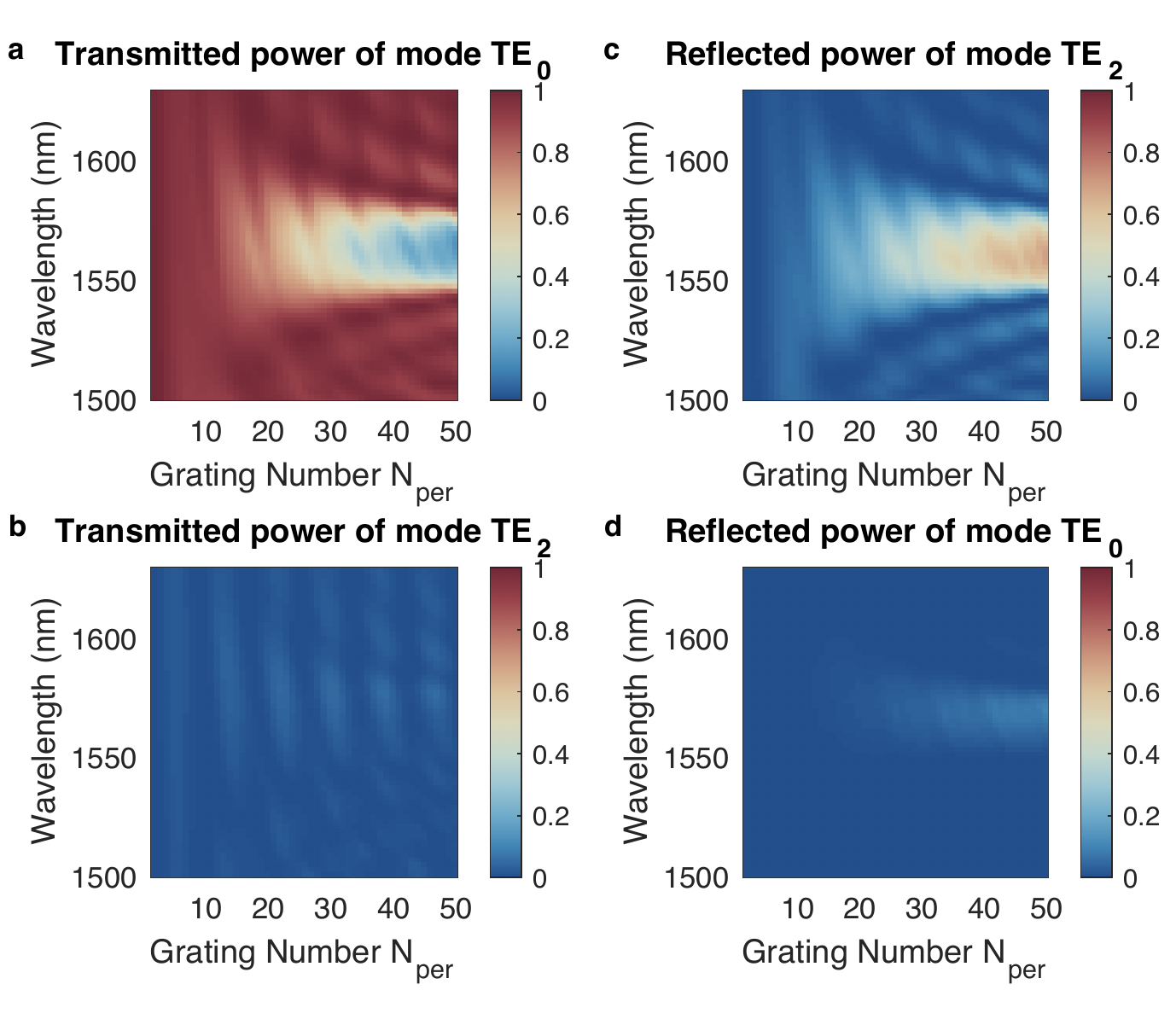} \\
\caption{\textbf{Contra-directional mode-converting grating properties under an incident $\mathrm{TE_0}$ mode.} \textbf{a-b,} Transmitted power into mode $\mathrm{TE_0}$ and $\mathrm{TE_2}$. (\textbf{c-d,}) Reflected power into mode $\mathrm{TE_0}$ and $\mathrm{TE_2}$.  Overall, these graphs demonstrate an efficient and selective conversion from input mode  $\mathrm{TE_0}$ into an output mode with transverse profile $\mathrm{TE_2}$, and that this conversion is efficient in a reflection geometry around 1560~nm. Furthermore, the Bragg grating reflects back into the same mode $\mathrm{TE_0}$ only minimally, and also the co-directional conversion is negligible. Furthermore, as the number of grating periods increases, the reflected power also increases. We chose a number of periods $\mathrm{N_{per}= 36}$.}
\label{fig_NperK1IN}
\end{figure}

In the Fig.~\ref{fig_NperK1IN}, we report the simulated reflection and transmission properties of the mode converting grating with parameters as described above, under an incident $\mathrm{TE_0}$ mode. We find that an efficient mode conversion is provided by the grating around 1560~nm, and that reflection into the same mode is negligible. Furthermore, with an increasing number of grating periods, the reflected power at the grating increases. 

\begin{figure}[tbh]
\centering
\includegraphics[width=14 cm]{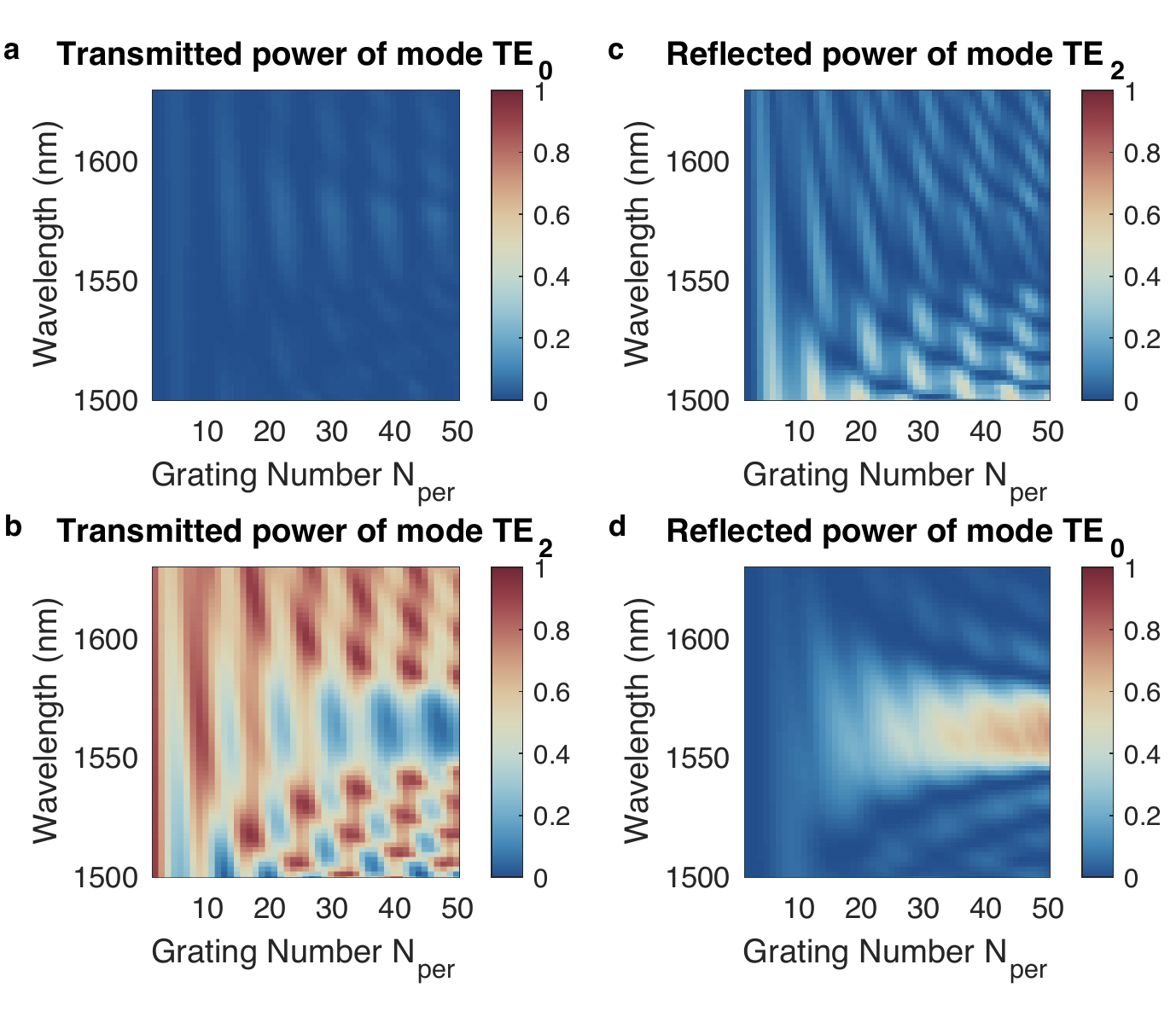} \\
\caption{\textbf{Contra-directional mode converting grating properties under an incident $\mathrm{TE_2}$ mode.} \textbf{a-b,} Transmitted power into mode $\mathrm{TE_0}$ and $\mathrm{TE_2}$. \textbf{c-d,} Reflected power into mode $\mathrm{TE_0}$ and $\mathrm{TE_2}$.  Overall, these graphs demonstrate an efficient and selective conversion from input mode  $\mathrm{TE_2}$ into the output mode with transverse profile $\mathrm{TE_0}$, and that this conversion is efficient in a reflection geometry around 1560~nm. Furthermore, the Bragg grating  reflects back into the same mode $\mathrm{TE_2}$ only minimally, and also the co-directional conversion is negligible. Furthermore, as the number of grating periods increases, the reflected power also increases. We chose a number of periods $\mathrm{N_{per}= 36}$.}
\label{fig_NperK3IN}
\end{figure}

In the Fig.~\ref{fig_NperK3IN}, we report the simulated reflection and transmission properties of the mode converting grating with parameters as described above, under an incident $\mathrm{TE_2}$ mode. We find that an efficient mode conversion is provided by the grating around 1560~nm, and that reflection into the same mode is negligible. Furthermore, with an increasing number of grating periods, the reflected power at the grating increases. By comparing panel~c of Fig.~\ref{fig_NperK1IN} with panel~d of Fig.~\ref{fig_NperK3IN}, we observe the reciprocal behavior of the mode converters. 

\begin{figure}[tbh]
\centering
\includegraphics[width=14 cm]{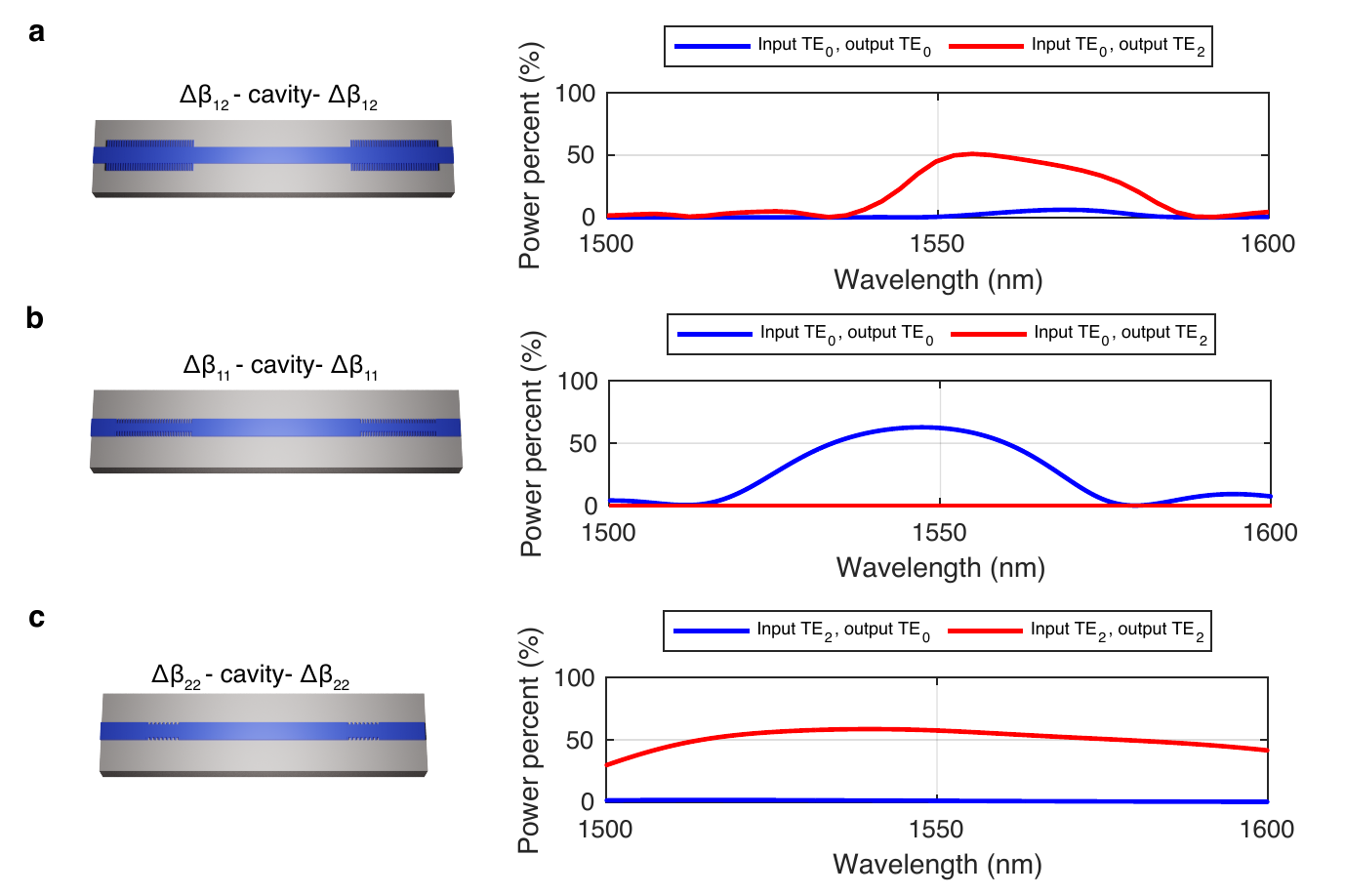} \\
\caption{\textbf{Reflection properties of mode-converting and non-mode converting gratings.} \textbf{a-c,} The reflection curves of the various Bragg gratings used in the resonators shown in the left panels are reported under different input and output conditions. } 
\label{fig_SOIproperties_3}
\end{figure}

Finally, we report in Fig.~\ref{fig_SOIproperties_3} the reflection curves for all types of Bragg grating, mode-converting~(a) and standard non-mode converting~(b,c). 

\subsection{Cascaded-mode resonators and Fabry-Perot resonators}

\begin{figure}[t!]
\centering
\includegraphics[width=18 cm]{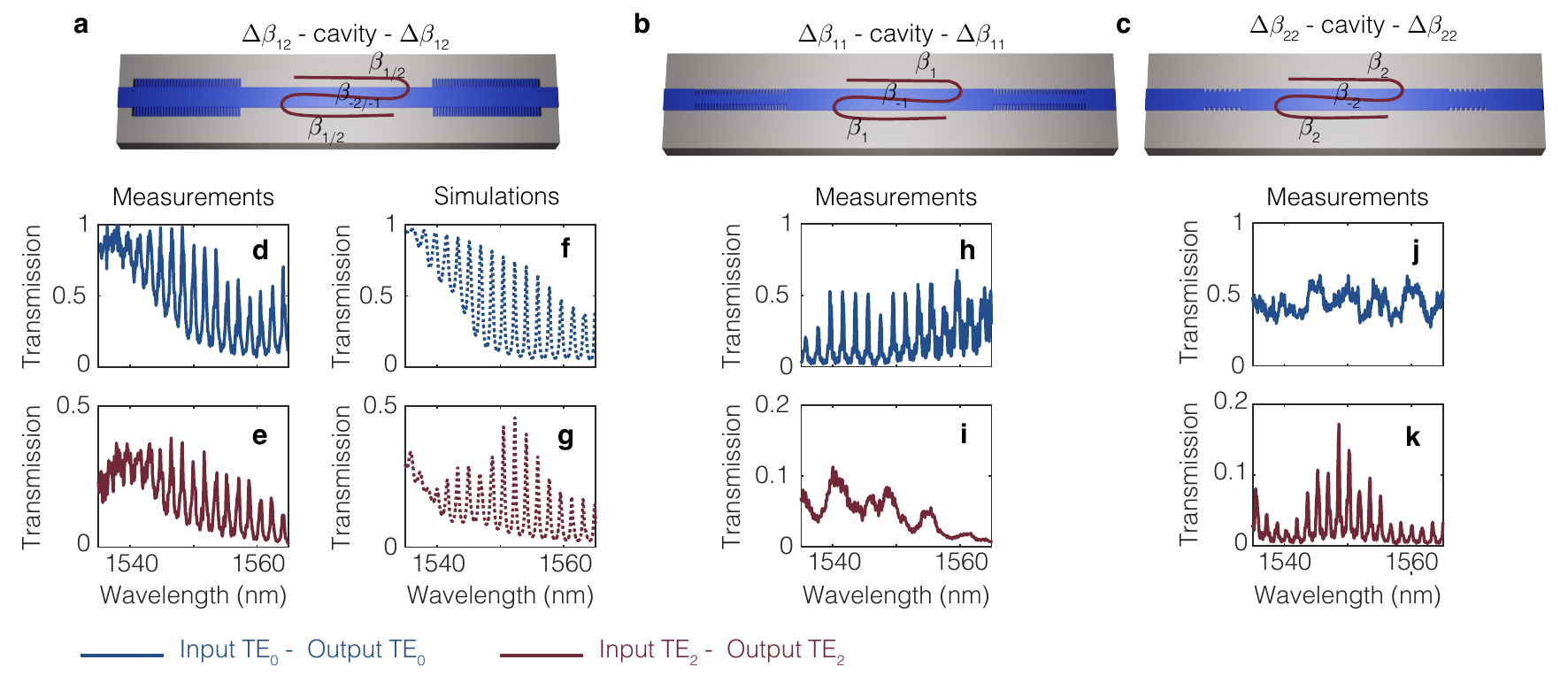} \\
\caption{\textbf{Transmission spectroscopy of cascaded-mode resonators versus conventional Fabry-Perot resonators.} \textbf{a,} A cascaded-mode resonator, where a reflection at both the left and the right Bragg mirror results into a conversion of the transverse mode from $\mathrm{TE}_0$ to $\mathrm{TE}_2$ and vice-versa, is compared to conventional Fabry-Perot resonators, where no mode conversion occurs upon reflection. Two types of Fabry-Perot resonators are considered, where the Bragg mirror provides selective reflection to either mode $\mathrm{TE}_0$ \textbf{b,} or mode $\mathrm{TE}_2$ \textbf{c}. \textbf{d-e,} The measured transmission spectra of the cascaded-mode resonator exhibit resonances regardless of whether $\mathrm{TE}_0$ or $\mathrm{TE}_2$ is incident onto the resonator. These  mode-independent resonances are a defining feature of cascaded-mode resonators. \textbf{f-g,} Simulated transmission spectra reproduce well the measurements. \textbf{h-i,} Measured transmission of a Fabry-Perot resonator for the case that either $\mathrm{TE}_0$ or $\mathrm{TE}_2$ is incident onto the resonator. The Bragg mirrors reflect selectively only $\mathrm{TE}_0$ and no mode conversion occurs. Cavity modes appear only for the $\mathrm{TE}_0$ input. \textbf{j-k,} In this case, the Bragg mirrors only reflect $\mathrm{TE}_2$ and no mode conversion occurs. Cavity modes appear only for the $\mathrm{TE}_2$ input. In all measurements and simulations, the analyzed mode (output mode) is the same as the probe mode (input mode) and the transmission curves are normalized to the transmitted intensity of a bare waveguide without any resonator. Blue curves are transmission measurements under a $\mathrm{TE}_0$ probe/analyzed mode, whereas red curves are transmission measurements under a $\mathrm{TE}_2$ probe/analyzed mode.}
\label{fig_modeparameters}
\end{figure}

Cascaded-mode resonators as discussed in the main text provide resonant confinement to input modes that correspond to either $\mathrm{TE}_0$ or $\mathrm{TE}_2$ transverse modes and have the same transmission spectrum for either input. We contrast here this property with two test Fabry-Perot resonators that employ standard mirrors and provide cavity confinement to only one of $\mathrm{TE}_0$ or $\mathrm{TE}_2$ modes. The experimental results are shown for the three cases in Fig.~\ref{fig_modeparameters}: cascaded-mode resonator (measurements d-e, simulations f-g), Fabry-Perot resonator operating on the $\mathrm{TE}_0$ mode (h-i) and Fabry-Perot resonator operating on the $\mathrm{TE}_2$ mode (j-k). By comparing the three cases, we find that cavity modes appear, as expected, for both $\mathrm{TE}_0$ and $\mathrm{TE}_2$ modes in the case of the cascaded-mode resonator only. Moreover, the experimental results are well-reproduced by our simulations. Cavity modes appear only for one of the two modes for the conventional Fabry-Perot resonators, while light is simply transmitted for the other modes.

\newgeometry{textwidth=18cm}

